\documentclass[alpha-refs]{wiley-article} 

\usepackage{color}
\usepackage[normalem]{ulem}
\usepackage{ifthen}
\usepackage{amsmath}
\usepackage{amssymb}
\usepackage{wasysym}
\usepackage{pifont}
\usepackage{graphicx,overpic}
\usepackage{mathrsfs}
\usepackage{caption}
\usepackage{multirow}
\usepackage{soul,xcolor}

\newcommand{\eqnref}[1]{{\ref{eq:#1}}}
\newcommand{\figref}[1]{\ref{fig:#1}}
\newcommand{\tabref}[1]{\ref{tab:#1}}
\newcommand{\pe}{P_\text{e}}
\newcommand{\ve}[1]{\ensuremath{\mbox{\boldmath$#1$}}}
\newcommand{\das}{\text{Da}_\text{s}}
\newcommand{\dad}{\text{Da}_\text{d}}
\renewcommand{\k}{\ensuremath{\text{\small TKE}}}
\newcommand{\ratio}{\mathscr{R}}
\newcommand{\noise}{\eta}
\newcommand{\chio}{\chi_0}
\newcommand{\airdensity}{\varrho_{\rm a}}

\definecolor{matlabgreen}{RGB}{119,172,48}
\definecolor{matlabred}{RGB}{217,83,25}
\definecolor{matlabblue}{RGB}{0,114,189}
\definecolor{matlabpurple}{RGB}{126,47,142}
\definecolor{matlabyellow}{RGB}{237,177,32}
\definecolor{matlablightblue}{RGB}{0,114,189}
\newcommand{\eng}[2]{{#2}}

\newcommand{\Dimensionless}{Non-dimensional}
\newcommand{\dedimensionalise}{non-dimensionalise}

\newcommand{\dimensionless}{non-dimensional}
\newcommand{\dimensionfull}{dimensional}
\newcommand{\Dimensionfull}{Dimensional}

\setstcolor{red}

\papertype{Research Article}
\title{Key parameters for droplet evaporation and mixing at the cloud edge}
\author[1]{J. Fries}
\author[2]{G. Sardina}
\author[3]{G. Svensson}
\author[1]{B. Mehlig}
\affil[1]{Department of Physics, Gothenburg University, SE-41296 Gothenburg, Sweden}
\affil[2]{Department of Mechanics and Maritime Sciences, Division of Fluid Dynamics, Chalmers University of Technology, SE-41296 Gothenburg, Sweden}
\affil[3]{Department of Meteorology, Stockholm University and Swedish e-science Research Centre, Stockholm, Sweden}
\corraddress{Bernhard Mehlig, Department of Physics, Gothenburg University, SE-41296 Gothenburg, Sweden}
\corremail{bernhard.mehlig@physics.gu.se}
\fundinginfo{Vetenskapsrådet, grant number: 2017-3865; Formas, grant number: 2014-585; Knut and Alice Wallenberg Foundation, Dnr. KAW 2014.0048}
\runningauthor{J. Fries et al.}

\begin{document}

\maketitle

\begin{abstract}
The distribution of liquid water in ice-free clouds determines their radiative properties, a significant source of uncertainty in weather and climate models.  Evaporation and turbulent mixing cause a cloud to display large variations in droplet-number density, but quite small variations in droplet size \citep{beals2015holographic}. Yet direct numerical simulations of the joint effect of evaporation and mixing near the cloud edge predict quite different behaviors, and it remains an open question how to reconcile these results with the experimental findings. To  infer the history of mixing and evaporation from observational snapshots of droplets in clouds is challenging  because clouds are transient systems.  We formulated a  statistical model that provides a reliable description of the eva\-po\-ra\-tion\--mixing process as seen in direct numerical simulations, and allows to infer important aspects of the history  of  observed droplet populations, highlighting the key mechanisms at work, and explaining the differences between observations and simulations. 

\end{abstract}

\section{Introduction}
Clouds play a major role in regulating weather and climate on Earth, by modulating the incoming solar radiation. Despite substantial scientific advances in the last decades, clouds  still represent the primary source of uncertainty in climate projections \citep{pincus2018shallow,IPCC}.
A key challenge is to understand how entrainment of dry air at the edges of ice-free clouds affects the size distribution and number density of droplets \citep{Blyth1993}.
This is important  because amount and distribution of liquid water determine cloud optical properties \citep{Kokhanovsky2004} and precipitation efficiency \citep{Burnet2007}. As a consequence, weather and climate models are sensitive to how entrainment  at the cloud edge is \eng{parameterized}{parameterised} \citep{mauritsen2012tuning}.

The optical properties of clouds are of crucial importance for the radiation balance of the Earth's climate system \citep{caldwell2016quantifying,dufresne2008assessment}. 
The size distribution and number density of droplets are key ingredients, because the light-extinction coefficient of the cloud is determined by the 
number of the droplets it contains, times their average surface area \citep{Kokhanovsky2004}. 

Regarding precipitation efficiency, the mechanism of rain formation in ice-free clouds is a longstanding unresolved problem in atmospheric physics \citep{Grabowski2013}.
A broad initial droplet-size distribution is needed to activate the collisions and coalescences of droplets that are necessary for the rapid onset of rain formation 
observed empirically in warm clouds \citep{devenish2012droplet,szumowski1997microphysical,szumowski1998microphysical}. 
Microscopic droplets grow by condensation of water \eng{vapor}{vapour}, or shrink by evaporation.
Yet when a droplet-containing parcel does not mix with its surroundings, condensation causes the droplet-size distribution to narrow because the diffusional growth of a droplet is inversely proportional to its radius, so that small droplets grow faster than large ones \citep{rogers1989short}.

Turbulence has a strong influence upon droplet condensation and evaporation in clouds \citep{bodenschatz2010can}.
Turbulent mixing causes water-\eng{vapor}{vapour} and liquid-water content to fluctuate on different length and time scales \citep{Vaillancourt:2001}. As a consequence, nearby droplets may have experienced quite different growth histories.
The droplets of a cloudy parcel that is mixed with dry air evaporate at different rates, so that the droplet-size distribution broadens \citep{Lanotte:2009,sardina2015continuous,sardina2018broadening,li2020condensational}. 
Cloud-resolving simulations \citep{hoffmann2019entrainment} show that this mechanism can have
a strong effect on droplet-number densities in turbulent clouds.

\begin{figure}[t]
\centering
\includegraphics{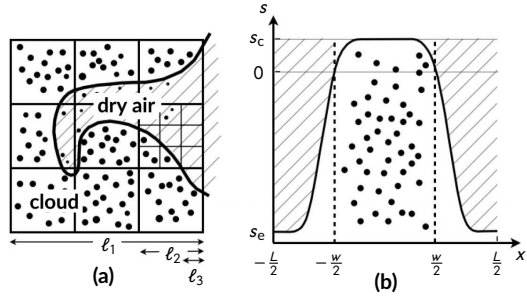}
\caption{\label{fig:schematic} 
({\bf a}) Effect of mixing on local droplet populations (schematic).
On large spatial scales ($\ell_1$ and $\ell_2$) mixing and evaporation is not yet complete, but some smaller regions
(of size $\ell_3$) are in local steady states: saturated air with droplets, or subsaturated air without droplets.
({\bf b}) Initial cloud configuration in our model. Before mixing, moist air and droplets reside in a $w \times L\times L$ slab, contained in a  cubic domain of side length $L$. Regions with dry  air are dashed. The solid line is the initial profile of supersaturation $s$ (see text).
}
\end{figure}

Entrainment of dry air at the cloud edge triggers rapid changes in the droplet-size distribution \citep{perrin2015lagrangian,abade2018broadening}.
As turbulence mixes dry air into the cloud it creates long-lasting regions of dry air where droplets can rapidly evaporate.
Droplets in regions with higher water-\eng{vapor}{vapour} concentration, by contrast, may saturate the air and survive for a much longer time, Figure~\figref{schematic}({\bf a}).
While droplets evaporate turbulence mixes the cloud at many length scales, ranging from
the Kolomogorov length -- of the order of \eng{millimeters}{millimetres} \citep{devenish2012droplet} -- to a few \eng{kilometers}{kilometres} \citep{rogers1989short}.
Evaporation and mixing on a spatial scale $\ell$ depend on the  turbulent mixing time $\tau_\ell$ at that scale, and upon the relevant thermodynamic time scale $\tau$.
Their ratio forms a Damk\"ohler number ${\rm Da} = \tau_\ell/\tau$ \citep{dimotakis2005turbulent}.
The thermodynamic process \eng{parameterized}{parameterised} by ${\rm Da}$ is limited by the rate of mixing if ${\rm Da}$ is large, and limited by thermodynamics if ${\rm Da}$ is small.
The dynamics at large Damk\"ohler numbers is  referred to as inhomogeneous mixing \citep{Baker1980}, 
where some droplets evaporate completely while others do not evaporate at all.
Small-Da mixing is called homogeneous, where droplets evaporate at approximately the same rate, so that the droplet-size distribution remains narrow. 

The notion of homogeneous and inhomogeneous mixing remains debated \citep{tolle2014effects}, but it can be given a precise meaning in terms of the fraction $\pe(t)$ of droplets that have completely evaporated at time $t$. 
However, it is at present not understood which mechanisms and parameters that determine the transition from homogeneous to inhomogeneous mixing.
Several authors have attempted to describe turbulent mixing in terms of one Damk\"ohler number. 
\cite{lehmann2009homogeneous} used a combined microphysical response time $\tau_{\rm r}$, a function of the two thermodynamic time scales of the problem, $\tau_{\rm d}$ (droplet evaporation) and $\tau_{\rm s}$ (supersaturation relaxation). 
\cite{lu2018microphysical}, by contrast, suggest that $\tau_{\rm d}$ should be used to formulate a single-parameter criterion for inhomogeneous mixing.
The direct numerical simulations (DNS)  by \cite{kumar2018scale} explored how the nature of mixing changes as  $\tau_{L}/\tau_{\rm r}$  increases
with the linear size $L$ of the simulated domain.
However, no clear sign of inhomogeneous mixing was found.
The authors mention that this may be a consequence of the thermodynamic setup used. Another possibility is that the simulated system was just not large enough.
\cite{Jeffery2007}  \eng{emphasized}{emphasised} that evaporation and mixing cannot be described by a single Da alone, because there are two thermodynamic time scales,
leading to three key \dimensionless~parameters,  ${\rm Da}_{\rm d}$, ${\rm Da}_{\rm s}$, and  the 
volume fraction $\chi$ of cloudy air.
However, Jeffery only studied the case ${\rm Da}_{\rm d}={\rm Da}_{\rm s}$ and did not discuss the implications of varying the Damk\"ohler numbers separately.
\cite{pinsky2016theoretical3} and \cite{pinsky2018theoretical4} modified an equation for advected liquid water \citep{rogers1989short} into a diffusion-reaction equation for the droplet-size distribution, and \eng{emphasized}{emphasised} the significance of a \dimensionless~parameter, their potential evaporation parameter.

Here we derive a statistical model for evaporation and turbulent mixing at the cloud edge
from first principles.
The model takes into account the multi-scale turbulent dynamics, as turbulent clouds can have large Reynolds numbers; ${\rm Re}\sim 10^7$ is a conservative estimate for convective clouds \citep{devenish2012droplet}.
The model quantitatively  predicts 
the outcomes of the DNS by \cite{kumar2012extreme,Kumar:2013,kumar2014lagrangian,kumar2018scale}.
Furthermore, the model shows in accordance with \cite{Jeffery2007} that the evolution of the droplet-size distribution is determined by ${\rm Da}_{\rm d}$, ${\rm Da}_{\rm s}$, and $\chi$. 
We find that the potential evaporation parameter of \cite{pinsky2016theoretical3} and \cite{pinsky2018theoretical4} is simply determined by the ratio of the Damk\"ohler numbers.
Finally, the model allows to interpret the results of in-situ measurements of clouds at the \eng{centimeter}{centimetre} scale \citep{beals2015holographic}.\\[0.1cm]

\section{Method}
We study mixing and evaporation of cloud droplets by mixing moist air, droplets, and dry air in a cubic domain of side length $L$
with periodic boundary conditions.
Initially, the saturated or slightly supersaturated moist air with supersaturation $s_{\rm c} \geq 0$ is contained in a $w\times L \times L$ slab together with $N_0$ randomly distributed water droplets, Figure~\figref{schematic}({\bf b}).
The dry air, initially outside the slab,  has negative supersaturation $s_{\rm e} < 0$. The mixing is driven by 
statistically stationary homogeneous isotropic turbulence, with turbulent kinetic energy $\k$
and mean dissipation rate per unit mass $\varepsilon$ \citep{frisch_1995}.
Essentially the same setup is  used in the DNS of \cite{kumar2012extreme,Kumar:2013,kumar2014lagrangian,kumar2018scale},
which allows us to understand their simulation results in terms of our model.

\subsection{Microscopic equations}
For the turbulent mixing, we start from the microscopic equations of \cite{kumar2018scale} and earlier studies \citep{vaillancourt2002microscopic,paoli2009turbulent,Lanotte:2009,kumar2014lagrangian,perrin2015lagrangian}.
We neglect buoyancy, particle inertia and settling, temperature changes due to vertical motion, temperature and pressure dependencies of the thermodynamic coefficients, and subsume the joint effects of temperature and water \eng{vapor}{vapour} into a single supersaturation field.

We denote fluid velocity and pressure by $\ve{u}(\ve{x},t)$ and $p(\ve{x},t)$, and supersaturation by $s(\ve{x},t)$.
The spatial position of droplet $\alpha$ is $\ve x_\alpha(t)$, its radius equals  $r_\alpha(t)$, and the index $\alpha$ ranges from $1$ to $N_0$.
We \dedimensionalise~as follows: 
$\ve{u}' = \ve{u}/U$, $\ve{x}' = \ve{x}/(U\tau_L)$, $t' = t/\tau_L$, $p' = p/(\airdensity U^2)$, $\ve{x}_\alpha' = \ve{x}_\alpha/(U\tau_L)$, $s' = s/|s_{\rm e}|$, $r_\alpha' = r_\alpha/r_0$, and  $s_{\rm c}' = s_{\rm c}/|s_{\rm e}|$. Here 
 $U = \sqrt{2\,\k/3}$ is the turbulent r.m.s. velocity, $\tau_L = \k/\varepsilon$ is the large-eddy time [$\propto L/U$ if the size of the largest eddies is of the order $L$ \citep{pope2000turbulent}], $\airdensity$ is the reference mass density  of air, $r_0 = [N_0^{-1}\sum_{\alpha = 1}^{N_0} r_\alpha(0)^3]^{1/3}$ is the initial volume-averaged droplet  radius,  and $|s_{\rm e}|$ is the (positive) subsaturation of the air outside the initial cloud slab.
Dropping the primes, the microscopic equations take the \dimensionless~form:
\begin{subequations}
\label{eq:model}
\begin{eqnarray}
&&\hspace*{-5mm}
\tfrac{{\rm D}}{{\rm D}t}\ve u
= - \nabla p +\text{Re}_L^{-1} \nabla^2 \ve{u} \label{veldl} \,\,\,\mbox{with}\,\,\,\nabla\cdot \ve{u}\! = \!0\,, \label{eq:veldlred} 
\\
&&\hspace*{-5mm}
\tfrac{{\rm D}}{{\rm D}t}s
 = ({\rm Re}_L {\rm Sc})^{-1}\nabla^2 s - \das \,\chi V \overline{r_\alpha(t) s(\ve x_\alpha,t)}  \,,\label{eq:supsatdlred}
\\
&&\hspace*{-5mm}
\tfrac{{\rm d}}{{\rm d}t}{\ve x}_\alpha=  \ve{u}(\ve x_\alpha,t) \,,\label{eq:droplmovdlred}\\
&&\hspace*{-5mm}
\tfrac{{\rm d}}{{\rm d}t}{r}_\alpha= 
\dad \,s(\ve x_\alpha(t),t) /(2r_\alpha) \quad\mbox{if $r_\alpha> 0 $}\,. \label{eq:droplraddlred}
\end{eqnarray}
\end{subequations}
Equations~\eqnref{veldlred} are the incompressible Navier-Stokes equations, with Lagrangian time derivative $\tfrac{{\rm D}}{{\rm D}t}=\partial_t \!+\! (\ve{u}\cdot \nabla) $.
In DNS of Equations~\eqnref{model}, a forcing is imposed to sustain stationary turbulence. This is not necessary in the model introduced below, and we therefore do not include a forcing term in Equation~\eqnref{veldlred}.
Equation~\eqnref{supsatdlred} is the equation for supersaturation.
The first term on its r.h.s.  describes diffusion of the supersaturation $s(\ve x,t)$. The second term models the effect of condensation and evaporation through the average $\overline{r_\alpha (t)s(\ve x_\alpha,t)}$,  taken over all droplets in the vicinity of~$\ve x$.
Droplets are advected by the turbulent flow (Equation~\eqnref{droplmovdlred}), and Equation~\eqnref{droplraddlred} models how the droplet radius $r_\alpha$ changes due to evaporation and condensation.
When a droplet has evaporated completely, we impose that it must remain at  $r_\alpha=0$.
In the derivation of Equation~\eqnref{droplraddlred}, $s(\ve x_\alpha(t),t)$ enters as the supersaturation at distances from the droplet that are much larger than the droplets radius  \citep{rogers1989short}.
In other words, Equation~\eqnref{droplraddlred} relies on a scale separation between droplet sizes and lengths that characterise supersaturation fluctuations generated by turbulent mixing \citep{Vaillancourt:2001}.
As a consequence, droplets interact locally with the supersaturation field over finite volumes through the average $\overline{r_\alpha (t)s(\ve x_\alpha,t)}$ in Equation~\eqnref{supsatdlred}.
Further details regarding Equations~\eqnref{model} are given in the Supporting Information (SI), where we also show how to derive Equations~\eqnref{model} from the more detailed dynamical description of \cite{Vaillancourt:2001,vaillancourt2002microscopic}, \cite{kumar2014lagrangian,kumar2018scale}, and \cite{perrin2015lagrangian}.

An advantage of writing the dynamics in \dimensionless~form is that this determines the independent \dimensionless~parameters. First, ${\rm Re}_L=\tfrac{2}{3} \k^2/(\varepsilon\nu)$ is the turbulence Reynolds number \citep{pope2000turbulent},  $\nu$ is the kinematic viscosity of air. 
The Schmidt number is defined as $\text{Sc} = \nu/\kappa$, where $\kappa$ is the diffusivity of~$s$.
The volume fraction of cloudy air is given by $\chi = w/L$, and $V = [L/(U \tau_L)]^3$ is the dimensionless domain volume.
The Damk\"ohler number $\das$ is defined as $\das = \tau_L/\tau_{\rm s}$, where $\tau_\text{s} = (4 \pi A_2 A_3 \varrho_\text{w} n_0 r_0)^{-1}$ is the supersaturation relaxation time. This is the time scale at which the supersaturation decays towards saturation, assuming that all droplets have the same radius $r_0$, and for
 droplet number density $n_0= N_0/(wL^2)$. Further, $\varrho_\text{w}$ is the density of pure liquid water, and $A_2$ and $A_3$ are thermodynamic coefficients, specified in the SI.
The Damk\"ohler number $\dad$ is defined as $\dad = \tau_L/\tau_{\rm d}$, where $\tau_\text{d} = r_0^2/(2 A_3 |s_{\rm e}|)$ is the droplet evaporation time,  the time that it takes for a droplet of radius $r_0$ to evaporate completely in a constant ambient supersaturation $s_{\rm e} < 0$.

The Damk\"ohler numbers determine the extent to which saturation and droplet evaporation are limited by the rate of mixing \citep{dimotakis2005turbulent}.
Saturation is mixing  limited at large $\das$, since regions with evaporating droplets --  created by mixing of  cloudy and dry air  at the time scale $\tau_L$ --
saturate faster than $\tau_L$.
When $\das$ is small, by contrast, evaporating droplets saturate the air more slowly than it is mixed. In this case, saturation is not limited by the rate of mixing.
Droplet evaporation is mixing limited at large $\dad$, since droplets then evaporate more rapidly than the exposure to subsaturated air changes.
At small $\dad$, mixing is faster than droplet evaporation, and droplets tend to evaporate mainly after the system has been mixed. The droplets then experience roughly the same supersaturation as they evaporate.

In the limit ${\rm Re}_L \rightarrow \infty$, three key \dimensionless~parameters remain in Equations~\eqnref{model}: $\chi$, $\dad$, and $\das$.
The system can be parameterised by $\chi$, $\dad$, and
\begin{equation}
\label{eq:ratio}
\ratio=\dad/\das\,.
\end{equation}
In this way, the scale dependence of the mixing process is contained in $\dad$ only.
The Damk\"ohler-number ratio $\ratio$ is inversely proportional to the density of liquid water in the cloud slab; it regulates the moisture of the mixing process (details in SI).
The bifurcation between moist steady states, where droplets remain in saturated air, and dry steady states, where all droplets have evaporated completely \citep{Jeffery2007,Kumar:2013,pinsky2016theoretical2}, occurs at a critical value of $\ratio$, $\ratio_{\rm c}$.
The critical ratio $\ratio_{\rm c}$ can be computed from the conserved quantity ${\theta = -\langle s(t)\rangle -\tfrac{2 \chi}{3 \ratio} [1-\pe(t) ]\langle r^3(t)\rangle}$, which is analogous to the liquid-water potential temperature at fixed altitude \citep{gerber2008entrainment,lamb2011physics,kumar2014lagrangian}.
Here, $\pe(t)$ is the fraction of completely evaporated droplets, the fraction of droplets for which $r_\alpha(t)=0$ at time $t$. Furthermore, $\langle s(t)\rangle = V^{-1}\int_V s(\ve x,t)\,\mathrm d \ve x$ is the volume average of supersaturation, and $\left \langle r^3(t) \right \rangle = \{[1-P_{\rm e}(t)]N_0\}^{-1}\sum_{\alpha = 1}^{N_0} r_\alpha(t)^3$ is the mean cubed droplet radius conditioned on $r_\alpha(t) > 0$ by the factor $[1-P_{\rm e}(t)]^{-1}$.
The conservation of $\theta$ can be concluded by integrating Equation~\eqnref{supsatdlred} for supersaturation over the domain volume, see the SI for details.
Moist steady states have $\left \langle r^3(t) \right \rangle > 0$ and $\langle s(t)\rangle = 0$, dry steady states have $P_{\rm e}(t) = 1$ and $\langle s(t)\rangle < 0$, so the sign of $\theta$ determines whether the steady-state is moist or dry.
The value of $\theta$ is determined by the initial conditions, $\theta = -\langle s(0)\rangle -{2 \chi}/({3 \ratio})$.
Setting $\theta = 0$, we find the critical Damk\"ohler-number ratio $\ratio_{\rm c} = -\tfrac{2}{3}\chi/\langle s(0)\rangle$.

DNS of Equations~\eqnref{model} for experimentally observed dissipation rates and droplet-number densities are feasible only for quite small systems \citep{kumar2018scale}.
This restricts the range of scales that can be explored, and makes it difficult to detect inhomogeneous mixing in DNS. We therefore pursue an alternative approach and 
adapt  a PDF model \citep{pope2000turbulent} -- commonly used to describe combustion processes \citep{haworth2010progress} -- to the inhomogeneous cloud edge.
As opposed to the kinematic statistical models reviewed
by \cite{Gus16}, we must also take thermodynamic processes into account.
\vfill\eject
\subsection{Statistical model}
Statistical models have been used to describe droplets in a supersaturation field that fluctuates around zero, as in the cloud core \citep{paoli2009turbulent,sardina2015continuous,chandrakar2016aerosol,siewert2017statistical}.  At the cloud edge, there are large deviations from this equilibrium. \cite{Jeffery2007}, \cite{pinsky2016theoretical3}, and \cite{pinsky2018theoretical4} formulated models for the cloud edge where  droplets evaporate in direct
response to a mean supersaturation field.
This does not take into account that mixing is local, and that small droplets are advected together with the supersaturation field.
For an accurate description of mixing and evaporation, it is essential to describe how each droplet carries its own local supersaturation \citep{siewert2017statistical}.
Our model does just that. It is derived from first principles using the established framework of  PDF models \citep{pope2000turbulent}.

For the configuration shown in Figure~\figref{schematic}({\bf b}) we derive one-dimensional statistical-model equations from Equations~\eqnref{model}
(details in the SI):
\begin{subequations}
\label{eq:statmod}
\begin{eqnarray}
&&\hspace*{-5mm}\mathrm d  u = -\tfrac{3}{4}C_0 u \mathrm d t  + \big( \tfrac{3}{2} C_0 \big)^{1/2} \mathrm d  \noise \,, \label{eq:velmc}\\
&&\hspace*{-5mm}\tfrac{\rm d}{{\rm d}t} s = - \tfrac{1}{2}C_\phi  [ s -\langle {s}(x,t)\rangle ] - \das\, \chi V \langle r(t) s( x,t)\rangle\,,
  \label{eq:supsatmc}\\
&&\hspace*{-5mm}\tfrac{\rm d}{{\rm d}t} {x} =  u \,, \label{eq:posmc}\\
&&\hspace*{-5mm}\tfrac{\rm d}{{\rm d}t} r =
\dad\, s /(2r)\quad \mbox{if}\quad  r > 0 \,. \label{eq:radmc}
\end{eqnarray}
\end{subequations}
Equation~\eqnref{velmc} describes the fluctuating acceleration of Lagrangian fluid elements in turbulence.
Here, $\mathrm d \noise$ are Brownian increments with zero mean and variance $\mathrm d t$, and $C_0$ is an empirical constant  \citep{pope2011simple}.
Each Lagrangian fluid element has a supersaturation $s$, and may contain a droplet (at position $x$, of size $r$).
Equation~\eqnref{supsatmc} approximates the supersaturation dynamics as decay towards $\langle s(x,t)\rangle$, regulated by the empirical constant $C_\phi$~\citep{pope2000turbulent}.
The second term on the r.h.s. of Equation~\eqnref{supsatmc} represents the effect of condensation and evaporation through $\langle r(t) s( x,t)\rangle$.
The position-dependent averages $\langle \cdots \rangle$ in Equation~\eqnref{supsatmc} are taken over fluid elements located at $x$ at time $t$ (details in the SI).
The statistical-model Equations \eqnref{statmod} becomes independent of $\text{Re}_L$ at large Reynolds numbers, where $C_0$ approaches a definite limit \citep{pope2011simple}. It is independent of $\text{Sc}$, in accordance with the known behaviour of advected scalars in fully developed turbulence \citep{shraiman2000scalar}.

Equations \eqnref{statmod} rest upon a probabilistic description of the dynamics of the two phases, droplets and air \citep{pope2000turbulent,jenny2012modeling}.
The corresponding evolution equations, dictated by Equations \eqnref{model}, contain unclosed terms that must be approximated \citep{pope1985pdf,haworth2010progress} to obtain a closed model such as Equations~\eqnref{statmod}.
Following \cite{pope2000turbulent}, we cast the model into the form of stochastic dynamical equations for Lagrangian fluid elements.
Since the dynamics is statistically one-dimensional in our configuration, we can average over the $y$ and $z$ coordinates to obtain Equations \eqnref{statmod} in one-dimensional form.
For the closures, we rely on standard approximations, common and justified in PDF modeling of single-phase flows \citep{pope1985pdf}, and in models for turbulent combustion \citep{haworth2010progress,jenny2012modeling,stollinger2013pdf}.
The explicit mathematical approximations for the closures provided in the SI render the interpretation of the statistical model definite, and indicate how to improve the model when necessary.

In the following, we briefly summarise the closures.
Equation \eqnref{velmc} contains the closure for fluid-element accelerations. It 
reproduces the empirically observed effect of turbulent diffusion of passive-scalar averages \citep{pope2000turbulent}.
Equation~\eqnref{supsatmc} contains two closure approximations.
First, the decay towards $\left \langle s(x,t)\right \rangle$ approximates the diffusion of supersaturation.
This closure ensures that the mean of a passive scalar is conserved, and that a passive scalar remains bounded between its minimal and maximal values \citep{pope2000turbulent}.
Furthermore, it describes the decay of passive-scalar variance in statistically homogeneous turbulent mixing of two scalar concentrations \citep{pope2000turbulent}.
However, this closure does not reproduce the relaxation of the single-point PDF of scalar concentration -- from a two-peaked distribution via a U-shaped distribution  into a Gaussian 
\citep{eswaran1988direct,pope1991mapping}.
For our initial conditions [Figure~\figref{schematic}{\bf (b)}], the decay towards $\left \langle s(x,t)\right \rangle$ captures the supersaturation fluctuations experienced by  a fluid element as  it moves towards or away from the most cloudy region.
Note that this closure does not account for  large saturated cloud structures that tend to relax slowly towards $\left \langle s(x,t)\right \rangle$.
Such events  are most relevant during the initial stages of the mixing-evaporation process, and their effect is expected to 
diminish with time, as large cloud structures are mixed into smaller and smaller structures.
Consequently, the statistical model may describe short initial transients only qualitatively, not quantitatively.

The second closure in Equation~\eqnref{supsatmc} approximates the effects of droplet phase change on the supersaturation field: the local average $\overline{r_\alpha (t)s(\ve x_\alpha,t)}$ in Equation \eqnref{supsatdlred} is replaced by the ensemble average $\left \langle r (t)s(x,t) \right \rangle$.
This is the simplest closure that preserves the conservation of the parameter $\theta$, and it is therefore common in PDF models that describe combustion of particles in turbulence \citep{haworth2010progress,jenny2012modeling,stollinger2013pdf}.
The average $\left \langle r (t)s(x,t) \right \rangle$  takes into account that droplet evaporation is delayed locally when nearby droplets  saturate the surrounding air.
Since we obtain closure by replacing the local average $\overline{r_\alpha (t)s(\ve x_\alpha,t)}$ by an average over fluid elements with one-dimensional dynamics,
 variations in the rate of supersaturation relaxation in the $y$ and $z$-coordinates  are not described.

It is expected that the  large cloud structures mentioned above, and their three-dimensional forms, matter more at very large Damk\"ohler numbers. Therefore it cannot be excluded that
the statistical model is only qualitative in this extreme limit. Below we show that the model works very well even for the largest Damk\"{o}hler numbers in DNS studies \cite{kumar2012extreme,kumar2014lagrangian}.
Also, since the statistical model is derived using the established framework of PDF models \citep{pope2000turbulent},  it can be straightforwardly improved by incorporating additional variables in the probabilistic description \citep{pope1990velocity,pope2000turbulent,meyer2008improved}, or by using more refined approximations \citep{pope1991mapping,jenny2012modeling}.
\\[0.1cm]

\begin{table}[bt]
\caption{\label{tab:simulations}
Summary of statistical-model simulations, details in SI. Damk\"{o}hler numbers $\dad$ and $\das$, Damk\"{o}hler-number ratio $\ratio$, critical ratio $\ratio_{\rm c}$, and volume fraction $\chi$ of cloudy air}.
\begin{threeparttable}
\begin{tabular}{llllll}
\headrow
\thead{Simulation} & $\dad$ & $\das$ & $\ratio$ & $\ratio_{\rm c}$ & $\chi$\\
\hiderowcolors
{Figure}~\figref{peoftime}~and~\figref{slab_comparison}({\bf a}) [dry]\mbox{}\hspace*{-3mm}  & 2.44   & 0.968  & 2.52     & 0.859     &    0.428      \\
{Figure}~\figref{peoftime} [moist]      & 1.09   &  1.43  &  0.76         & 0.859            & 0.428 \\
{Figure}~\figref{slab_comparison}({\bf b}) [very moist]          & 0.754   &  8.20   &  0.092        & 0.683        & 0.4 \\
{Figure}~\figref{paramplane} & 5E\raisebox{1mm}{-3}-4E\raisebox{1mm}{2} &  1E\raisebox{1mm}{-3}-9E\raisebox{1mm}{3} & 4E\raisebox{1mm}{-2}-4E\raisebox{1mm}{0}         & 0.913               & 0.429 \\
{Figure}~\figref{mapping}({\bf a})          & 1E\raisebox{1mm}{-2}-1E\raisebox{1mm}{3}& 6E\raisebox{1mm}{-2}-6E\raisebox{1mm}{3}  & 0.17               &     0.18-2.7          & 0.2-0.8    \\
{Figure}~\figref{mapping}({\bf b})          &  1E\raisebox{1mm}{-2}-8E\raisebox{1mm}{2} &  3E\raisebox{1mm}{-3}-4E\raisebox{1mm}{4}   & 2.4E\raisebox{1mm}{-2}-2.9E\raisebox{1mm}{-2 }         & 0.38-0.41           &  0.369-0.374 \\
\hline
\end{tabular}
\end{threeparttable}
\end{table}

\begin{table}[bt]
\setlength{\tabcolsep}{2.4pt}
\caption{\label{tab:otherstudies} Parameters of DNS shown in Figure~\ref{fig:paramplane}: Damk\"{o}hler number $\dad$, Damk\"{o}hler-number ratio $\ratio$, critical ratio $\ratio_{\rm c}$, and volume fraction $\chi$ of cloudy air. Some \dimensionfull~parameters are also shown: domain size $L$, mean dissipation rate $\varepsilon$, and droplet-number density $n_0$ of the initially cloudy air.}
\begin{threeparttable}
\begin{tabular}{>{}l>{}l>{}l>{}l>{}l>{}l>{}l>{}l>{}l}
\headrow
  &  \multicolumn{5}{c}{\bf \Dimensionless~parameters} & \multicolumn{3}{c}{\bf \Dimensionfull~parameters} \\[-1.0ex]
\headrow & & & & & & $\,\,\,L$&   $\,\,\,\,\,\,\,\varepsilon$ & $\,\,\,\,n_0$ \\[-0.9ex]
\headrow
\multirow{-1.5}{*}{\bf Reference}& \multirow{-1.5}{*}{$\dad$} & \multirow{-1.5}{*}{$\das$} &  \multirow{-1.5}{*}{$\ratio$}& \multirow{-1.5}{*}{$\ratio_{\rm c}$}  & \multirow{-1.5}{*}{$\chi$}  & [cm] &    [cm$^2$/s$^3$] &  [cm$^{-3}$]  \\
\hiderowcolors
  \multicolumn{1}{l|}{ $\circ$ \cite{Andrejczuk:2006}}&  \multicolumn{1}{|l}{ 8E$^{-1}$-1E$^2$} & 3E$^{0}$-3E$^2$ & 0.13-2.8 & 0.10-4.5 &  \multicolumn{1}{l|}{ 0.13-0.87} & \multicolumn{1}{|l}{ 64}  &  4E$^{-1}$-9E${^2}$ & 1E$^{2}$-1E${^3}$ \\
  \multicolumn{1}{l|}{$\triangle$ \cite{kumar2012extreme}}&   \multicolumn{1}{|l}{ 8E$^{-3}$-8E$^{-1}$} & 8E$^{-2}$-8E$^0$&  9.2E$^{-2}$ & 0.68 &  \multicolumn{1}{l|}{ 0.4} & \multicolumn{1}{|l}{ 26}  & 34 &  164  \\
  \multicolumn{1}{l|}{ $\square$ \cite{Kumar:2013}}&  \multicolumn{1}{|l}{0.14, 0.31} & 0.62, 0.41& 0.22, 0.73 & 0.68 &  \multicolumn{1}{l|}{0.4} & \multicolumn{1}{|l}{26}  & 34 & 164 \\
  \multicolumn{1}{l|}{ $\diamond$ \cite{kumar2014lagrangian}} &  \multicolumn{1}{|l}{ 0.61-2.4} & 0.97-1.9&  0.31-2.5 & 0.84 &  \multicolumn{1}{l|}{0.42} & \multicolumn{1}{|l}{ 51}  & 34 & 153 \\
  \multicolumn{1}{l|}{ $\triangledown$ \cite{kumar2018scale}}&  \multicolumn{1}{|l}{ 0.12-0.91} & 0.51-4.0 & 0.23  &0.90-0.95 &  \multicolumn{1}{l|}{ 0.42-0.45} & \multicolumn{1}{|l}{1E$^{1}$-2E${^2}$}  & 32-35 & 120  \\
\hline
\end{tabular}
\end{threeparttable}
\end{table}

\section{Results}

\subsection{Comparison with DNS}
\begin{figure*}[t]
\centering
\includegraphics{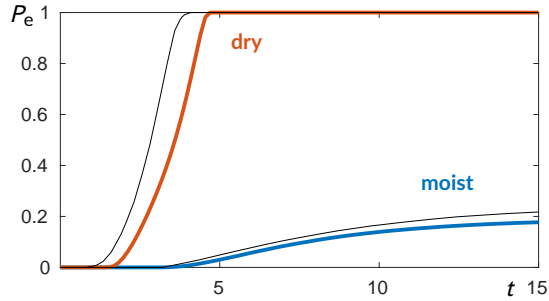}
\caption{\label{fig:peoftime}
Fraction $\pe(t)$  of droplets that have completely evaporated as a function of \dimensionless~time $t$,
parameters  in Table \tabref{simulations}.
Coloured lines are statistical-model simulations, black lines are DNS of \cite{kumar2014lagrangian}.}
\end{figure*}

\begin{figure*}[t]
\centering
\includegraphics{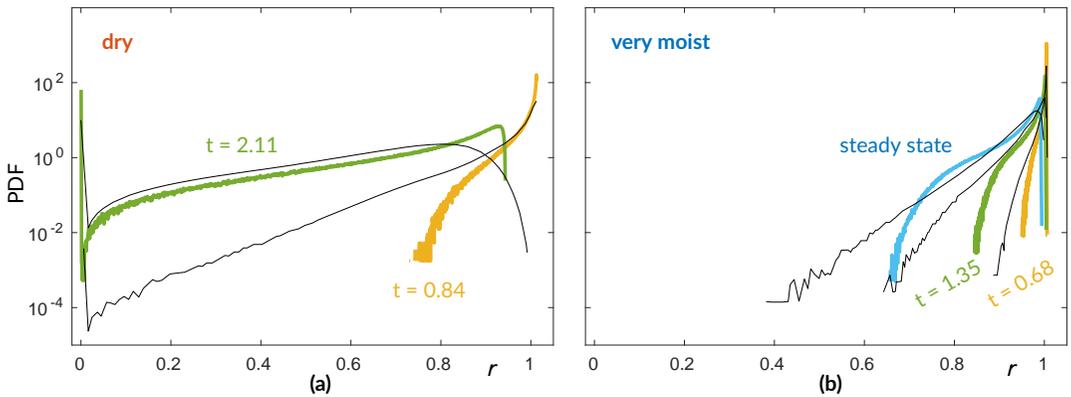}
\vspace{-2mm}
\caption{\label{fig:slab_comparison}
({\bf a}) Evolution of the droplet-size distribution (parameters  in Table~\tabref{simulations}) for different times.
The probability density of the \dimensionless~droplet radius $r$ is shown for the statistical-model (coloured lines) and DNS of \cite{kumar2014lagrangian} (black lines).
The initial droplet-size distribution is monodisperse, and centred at $r = 1$.
({\bf c}) Same as ({\bf b}), but for a very moist case (parameters  in Table~\tabref{simulations}) and DNS of \cite{kumar2012extreme}.}
\end{figure*}

The statistical model can be used to understand DNS results of 
\cite{kumar2012extreme,Kumar:2013,kumar2014lagrangian,kumar2018scale}.
Figure~\figref{peoftime} shows good agreement for the time evolution of the fraction $\pe(t)$
of droplets that have completely evaporated, even though the statistical-model dynamics is slightly slower. 
Panels ({\bf a}) and ({\bf b}) in Figure~\figref{slab_comparison} show that the model reproduces the broadening of the droplet-size distribution.
The slightly slower dynamics in Figure~\figref{peoftime} and the deviations in the tails in Figure~\figref{slab_comparison} suggest that the statistical model does not reproduce the most rapid evaporation rates.
This could be due to turbulent fluctuations in the supersaturation diffusion, neglected in Equation~\eqnref{supsatmc}.
\cite{kumar2012extreme,kumar2014lagrangian} compute droplet-size distributions with prominent exponential tails using DNS -- some of them are seen in Figure~\figref{slab_comparison} (black lines) -- and connect these tails to corresponding exponential tails in the PDF of supersaturation at droplet positions.
In our statistical-model simulations we observe heavy tails in the PDF of supersaturation at the droplet position, but the tails are less pronounced than in the DNS (not shown). Heavy tails are consistent with the results of \cite{eswaran1988direct} mentioned above,  who observed how an initially bimodal supersaturation relaxes.

Despite these shortcomings,  our model describes the time evolution of $\pe(t)$ very well (Figure~\figref{peoftime}). It is also a significant improvement over models in which
the droplets interact with a mean supersaturation field \citep{Jeffery2007,pinsky2016theoretical3,pinsky2018theoretical4}.  In reality, the droplets react to the local supersaturation, as mentioned above, and this may be particularly important at large values of $\dad$, where  locally saturated regions can persist for a long time.

\begin{figure}[h!]
\centering
\includegraphics{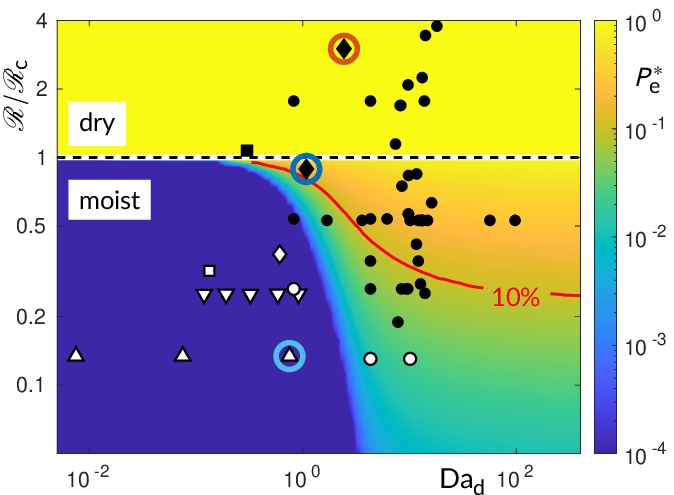}
\vspace{-2mm}
\caption{\label{fig:paramplane}
Steady-state fraction $\pe^*$  of completely evaporated droplets, as a function of $\dad$ and $\ratio/\ratio_{\rm c}$, details in SI. The fraction $\pe^\ast$ is \eng{color}{colour} coded. The solid red line is the contour $\pe^* = 10\%$,  the black dashed line indicates the transition between moist and dry steady states, and symbols indicate DNS results from previous studies: $\triangle$~\citep{kumar2012extreme}, $\square$~\citep{Kumar:2013}, $\diamond$~\citep{kumar2014lagrangian}, $\triangledown$~\citep{kumar2018scale}, and $\circ$~\citep{Andrejczuk:2006}. Filled symbols indicate $\pe^* > 10\%$. The DNS of \cite{Andrejczuk:2006} should not be quantitatively compared to the statistical model results, since they are for different initial conditions and decaying turbulence (see text and SI). Red, blue and light blue circles correspond to the dry, moist, and very moist simulations in Figures~\ref{fig:peoftime}~and~\ref{fig:slab_comparison}.}
\end{figure}

Figure~\figref{paramplane} shows the steady-state value $\pe^*$ of $\pe(t)$ computed from the statistical model as a function of $\dad$ and $\ratio/\ratio_{\rm c}$.
We see how $\pe^*$ increases with both $\dad$ and $\ratio/\ratio_{\rm c}$.
The DNS results of \cite{Andrejczuk:2006} and \cite{kumar2012extreme,Kumar:2013,kumar2014lagrangian,kumar2018scale} 
 form a pattern in Figure~\figref{paramplane} that verifies these dependencies: open symbols correspond to DNS with little or no complete 
 evaporation in the steady state ($\pe^*<10\%$), and filled symbols to $\pe^*>10\%$.
Figure~\figref{paramplane} also explains why the DNS of \cite{kumar2018scale} did not exhibit significant levels of inhomogeneous mixing: since their $\ratio$ was quite small, small values of $\pe^\ast$ require values of $\dad$ much larger than unity ($\dad \sim 10^2$ for  $\pe^\ast=10\%$).
Furthermore, the substantially different outcomes
of the DNS of \cite{kumar2012extreme} and \cite{kumar2014lagrangian} are explained, their parameters lie on opposite sides of the bifurcation line.
Figure~\figref{paramplane} also explains, at least qualitatively, numerical results of DNS of transient turbulence with quite different initial conditions~\citep{Andrejczuk:2006},
namely how the amount of complete droplet evaporation increases with both $\ratio/\ratio_{\rm c}$ and $\dad$.
There is no parameter corresponding directly to $\dad$ in the simulations of \cite{Andrejczuk:2006}, because they are for different initial conditions and flows.
We therefore place these simulations in Figure~\figref{paramplane} by computing a time-scale ratio that, in a qualitative sense, incorporates the same physics as $\dad$ (details in the SI).
Key parameters of the DNS in Figure~\figref{paramplane} are summarised in Table~\tabref{otherstudies}, a complete description is provided in the SI.

\eject

\subsection{Mixing histories from observations}
\label{mixhist}
\begin{figure}
\centering
\includegraphics{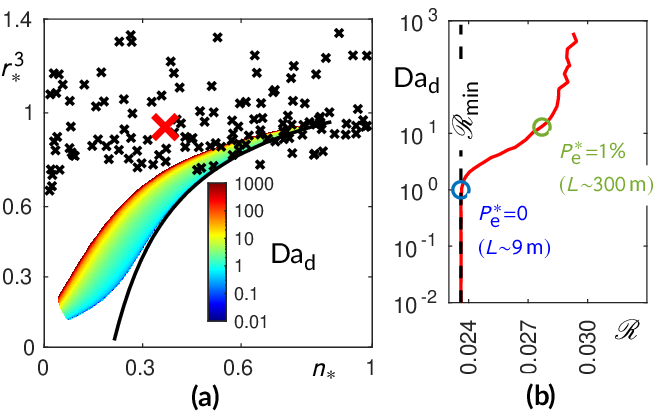}
\caption{\label{fig:mapping}
({\bf a}) Mixing diagram. Empirical data from \cite{beals2015holographic} (black crosses). 
The homogeneous mixing line (black) from top panel of Figure~3 of \cite{beals2015holographic} 
corresponds to $\ratio=0.17$.
The \eng{colored}{coloured} region shows where steady states are found in the mixing diagram for $\ratio=0.17$.
The corresponding values of $\dad$ are \eng{color}{colour} coded (legend).  
The red cross is the measurement shown in the middle panel of Figure~2 of \cite{beals2015holographic}. 
({\bf b}) Values of $\ratio$ and $\dad$ consistent with a steady state at the red cross in panel  ({\bf a}) (red).
We estimate $L\sim 9$ m for $\dad = 1$ (blue circle). These local mixing processes have $\ratio = \ratio_\text{min}$ (black dashed line), so no droplets evaporated completely ($\pe^\ast=0$). The green circle corresponds to $\dad = 13$ and $\ratio = 0.028$ with  $\pe^* = 1$\% and $L\sim 300\,$m.}
\end{figure}

A common way of characterising the droplet content of a cloud is to plot the mean cubed radius $r^3$ and number density $n$ of droplets for observed cloud-droplet populations in a mixing diagram.
Figure~\figref{mapping}({\bf a}) shows a mixing diagram with empirical data from \cite{beals2015holographic}.
Black crosses are values of $r^3$ and  $n$ extracted from snapshots (linear size $15\,$cm) of local droplet populations measured during an airplane flight through a convective cloud.

Observational data in mixing diagrams are commonly discussed in relation to the homogeneous mixing line, a curve of global steady states ($r_\ast^3,n_\ast$) that result from homogeneous mixing (no complete evaporation) between different proportions of undiluted cloudy and dry environmental air \citep{gerber2008entrainment,kumar2014lagrangian,pinsky2016theoretical3}.
\cite{beals2015holographic} calculated this line, it is also shown in Figure~\figref{mapping}({\bf a}).
A fundamental problem is however that it is not clear how to interpret mixing diagrams such as Figure~\figref{mapping}({\bf a}), since it is not clear that the empirically observed droplet populations reflect global steady states \citep{pinsky2018theoretical4}.

It is nevertheless likely that most data points in Figure~\figref{mapping}({\bf a}) sample local steady states, i.e. locally well-mixed droplet populations that reside in saturated air.
To describe such droplet populations, one must refer to the  multi-scale turbulent mixing process.
We attempted this analysis using the statistical model, assuming that the statistical model with the initial condition shown Figure~\figref{schematic}({\bf b}) describes how a cloud structure at the spatial scale $L$ develops.
Under this assumption, $n_\ast$ and $r^3_\ast$ are given by the droplet-number density (normalised by $n_0$) and the mean cubed droplet radius $\left \langle r(t)^3\right \rangle$ in the steady state, and we can conjecture the mixing histories that formed the measured droplet populations.

We begin by noting that  $\chi$ and $\pe^*$ are completely determined for any steady-state point ($r_\ast^3,n_\ast$) in a mixing diagram.
To show this, we write the volume-averaged initial supersaturation as $\langle s(0)\rangle = (1+s_{\rm c})(\chi + \chio)-1$, where $\chio$ is a constant that depends on the initial supersaturation profile (details in the SI). Inserting 
\begin{equation}
\label{eq:nast}
\chi = n_*/(1-\pe^*)
\end{equation}
 into $\theta = -\langle s(t)\rangle -\tfrac{2 \chi}{3 \ratio} [ 1-\pe(t) ]\langle r^3(t)\rangle$ we find:
\begin{align}\label{eq:r3}
\pe^* = 1 - \frac{n_\ast [1+\tfrac{3}{2} \ratio (1+s_{\rm c})]}
{n_\ast r_\ast^3+\tfrac{3}{2} \ratio [ 1-\chio(1+s_{\rm c})]}\, .
\end{align}
Equations~\eqnref{nast} and \eqnref{r3}  determine how to map ($r_\ast^3,n_\ast$) to  ($\chi,\pe^*$). 
As a consistency check we note that  one obtains the homogeneous mixing line \citep{pinsky2016theoretical3} from Equation~\eqnref{r3} by setting $\pe^* = 0$.
This allows us to infer that $\ratio = 0.17$ and $s_{\rm c} = \chio = 0$ from the homogeneous mixing line of \cite{beals2015holographic}.

Any point in the mixing diagram must correspond to a local steady state with certain values of $\pe^*$ and $\chi$.
Each statistical-model simulation for given $\dad$, $\ratio$, and $\chi$ yields a certain value of $\pe^*$. This
allows us to extract a value of $\dad$  for each point in the mixing diagram from our statistical-model simulations.
The result is shown in Figure~\figref{mapping}({\bf a}).
We see that $\dad$ increases  rapidly above the homogeneous mixing line.
Estimating $\tau_{\text{s}} \sim {1}\,$s from \cite{beals2015holographic} and conservatively estimating $\varepsilon \sim 1$ cm$^2$s$^{-3}$ for a convective cloud, 
a value of $\dad = 1000$ implies that an observed droplet population was mixed at spatial scales of the order of 
$L \sim \sqrt{\varepsilon\tau_L^3} \sim 5\,$km, larger than the size of the cloud. In other words, the rapid increase of $\dad$ in Figure~\figref{mapping}({\bf a}) suggests that most of the data in the mixing diagram cannot be in a global steady state of a mixing process  \eng{parameterized}{parameterised} by $\ratio=0.17$.

We concluded above that most measurements of \cite{beals2015holographic} are likely to correspond to local steady states.
As undiluted cloudy air is mixed with premixed  air, such steady states are formed locally and temporarily as local mixing processes equilibrate at small spatial scales [Figure~\figref{schematic}({\bf a})].
We now discuss how the analysis of local steady states may yield  insight into possible local histories of the cloud.
Air affected by earlier mixing events is not as dry as environmental air, so the mixing of undiluted cloud with premixed air is governed by smaller values of $\ratio$.
We therefore ask: which values of $\ratio$ are consistent with the assumption that the experimentally observed droplet population in the middle panel of Figure~2 of \cite{beals2015holographic} -- red cross in Figure~\figref{mapping}({\bf a}) -- reflects a local steady state?
Our model allows us to determine possible combinations of $\dad$, $\ratio$, and  $\chi$ consistent with a local steady state.
We know that $\ratio$ must be  smaller than 0.17, the upper limit dictated by the homogeneous
mixing line of  \cite{beals2015holographic}. Furthermore, since the data cannot lie below the homogeneous mixing line of the global mixing process, 
 a lower bound for $\ratio$ is $\ratio_{\rm min} = \tfrac{2}{3} (n_*-r^3_*) n_*/\left [(1+s_{\rm c})(\chio+n_*)-1\right ] = 0.0236$. 
 
 Figure~\figref{mapping}({\bf b}) shows values of $\ratio$ and $\dad$ obtained from our statistical-model simulations that
 are consistent with these constraints. We see  that the range of possible
 values of  $\dad$ covers several orders of magnitude. This means that local mixing processes consistent with the red cross in Figure~\figref{mapping}({\bf a}) may have occurred over a large range of spatial scales.
We also see that $\ratio$ does not vary much in Figure~\figref{mapping}({\bf b}), only between 0.024 and 0.03. This  allows us to conclude that some important aspects of the mixing dynamics are essentially independent of spatial scale.
First,  the fact that $\ratio$ is substantially smaller than 0.17 indicates that the non-cloudy air was premixed.
Second, using Equation~\eqnref{r3}, we find that the reduction in droplet-number density was primarily caused by dilution even at the largest scales,  since $\pe^*$ increases only up $\sim$1.4\% for the largest value of $\dad$, at
$\dad \sim 1000$ and $\ratio = 0.03$. Put differently, $\chi \sim n_\ast$ for all values of $\dad$ we considered.

How does the outcome of a local mixing process depend on its scale? Larger scales correspond to 
 larger values of $\dad$, and  Figure~\figref{mapping}({\bf b}) shows that complete droplet evaporation begins to occur around $\dad = 1$, where $\ratio$ starts to exceed $\ratio_\text{min}$ (blue circle). 
Estimating ${\varepsilon \sim 10}$~cm$^2$s$^{-3}$, a typical value for convective clouds \citep{devenish2012droplet}, 
we find that $\dad = 1$, $\tau_\text{s} \sim 1$, and $\ratio = \ratio_\text{min}$ correspond to the spatial scale $9\,$m.
Mixing processes leading to the red cross in Figure~\figref{mapping}({\bf a}) that occurred at scales smaller than  $9\,$m
 were therefore perfectly homogeneous: none of the droplets evaporated completely as they were diluted by premixed air.
At larger spatial scales, small but non-zero fractions of the droplets evaporated completely.
Equation~\eqnref{r3} gives $\ratio = 0.028$ for $\pe^* = 1\%$, and from Figure~\figref{mapping}({\bf b}) we read off $\dad = 13$~(green~circle).
For $\varepsilon \sim 10$~cm$^2$s$^{-3}$ these values correspond to $300\,$m.
This suggests that reductions in droplet number concentration are dominated by dilution, and not complete droplet evaporation, also for mixing processes that range over hundreds of meters.
Furthermore, since most data points in Figure~\figref{mapping}({\bf a}) reside well above the region where equilibria are found for $\ratio = 0.17$,  we conclude that they too resulted from mixing with premixed air. 

\section{Discussion}
A general conclusion from our analysis is that both Damk\"{o}hler numbers are important for the transition to inhomogeneous mixing; $\dad$ \eng{parameterizes}{parameterises} the mixing-limited nature of droplet evaporation, and 
the  ratio $\ratio= \dad/\das$ regulates the self-limiting effect of droplet evaporation, namely that droplets cease to evaporate when they have saturated the surrounding air or evaporated completely. \eng{Analyzing}{Analysing} the parameters of our microscopic equations \eqnref{model} we see that
$ \mathscr{R}=-\tfrac{3}{2} R$ where $R$ is the potential
evaporation parameter of \cite{pinsky2016theoretical3} and \cite{pinsky2018theoretical4}.
So $R$ is in fact given by the ratio of $\dad$ and $\das$, consistent with our conclusion that both
Damk\"ohler numbers matter.

\cite{pinsky2018edge1,pinsky2019edge2} concluded that the Damk\"{o}hler number ratio $\ratio$ determines whether a cloud expands by dilution or shrinks by complete droplet evaporation.
A mixing process that mixes equal proportions of saturated cloudy and subsaturated non-cloudy air has $\ratio_{\rm c} = \tfrac{2}{3}$, so the symmetric configuration they adopted for the cloud edge implies that the cloud expands if $\ratio < \tfrac{2}{3}$, and shrinks otherwise.
We note that whether the cloud expands or shrinks depends on the position and scale at which one perceives it.
A local mixing process with small $\ratio/\ratio_{\rm c}$ tends towards a moist steady state, so the cloud dilutes locally.
A local mixing process that tends towards a dry steady state, by contrast, consumes the cloud.
It is possible for a cloud to expand locally for some time, even if this local expansion is part of a mixing process that consumes the cloud at larger length and time scales.
Although global mixing processes are transient, they contain local steady states. Diluted and saturated local droplet populations [such as the red cross in {Figure~\figref{mapping}({\bf a})] can be a part of such transients. However, such local steady states must eventually be abandoned as mixing  proceeds globally.

Adopting this multi-scale picture of mixing, in which a large-scale mixing process consists of many mixing processes at smaller scales, it is natural to expect that large ranges of the parameters $\chi$, $\ratio$ and $\dad$ are relevant. If one moves the domain of a local mixing process from the interior of the cloud towards the cloud edge, the liquid water content decreases, so that $\ratio/\ratio_{\rm c}$ and $\dad$ increase.
\cite{lehmann2009homogeneous} point out that $\dad$  increases as one perceives mixing processes at larger and larger spatial scales. This corresponds to moving to the right in Figure~\figref{paramplane}. We note that  $\ratio/\ratio_{\rm c}$ and $\dad$ tend to increase with distance from the
interior of the cloud. 
Moving the sampling volume towards the cloud edge then corresponds to a motion upwards and to the right in Figure~\figref{paramplane}. The  amount of complete droplet evaporation increases in this direction, consistent with the fact that complete droplet evaporation takes place at the cloud edge. 

A number of assumptions may influence our interpretation of the empirical data in Section~\ref{mixhist}.
First, the model configuration in Figure~\figref{schematic}({\bf b}) is simplified compared to real clouds, which have irregular shapes that deform during the mixing-evaporation process.
Second, the observational method may not detect droplets with radii smaller than 3 $\mu$m ($r_\alpha^3 = 0.2$), as stated by \cite{beals2015holographic}.
If many small droplets where not detected, the observations in Figure~\figref{mapping}({\bf a}) are located too far from the homogeneous mixing line.
Third, at the upper end of the Damk\"{o}hler range in Figure~\figref{mapping}, the statistical model may not be quantitatively accurate, as stated above.
We nevertheless expect that the statistical model reproduces the evolution of droplet-size distributions in DNS qualitatively in Figure~\figref{mapping}. This expectation is corroborated by the robust tendency for $\pe^*$ to increase with increasing values of $\dad$ and $\ratio$ in Figure~\figref{paramplane}, and follows directly from the roles of the Damk\"{o}hler numbers in mixing-evaporation dynamics.

The deviations in the tails of droplet-size distributions at moderate Damk\"{o}hler numbers in Figure~\figref{slab_comparison} suggest that the next step in improving the statistical model should aim at reproducing the fastest evaporation rates in the transient mixing-evaporation process.
A better agreement in the tails could be achieved by refining the closure for supersaturation diffusion by using a dynamic $C_\phi$ in Equation~\eqnref{supsatmc} \citep{jenny2012modeling}, or by introducing additional fluctuations \citep{pope1991mapping}.
Another possibility is to refine the description of the spatial structure of the supersaturation field by 
improved  closures  \citep{vedula2001dynamics,pope1991mapping,meyer2008improved,jenny2012modeling}

\section{Conclusions}
We derived a statistical model for evaporation and turbulent mixing at the cloud edge from first principles. 
The model explains results of earlier DNS studies of mixing \citep{Andrejczuk:2006,kumar2012extreme,Kumar:2013,kumar2014lagrangian,kumar2018scale}, and shows that two thermodynamic time scales are important for a mixing process, the droplet evaporation time and the supersaturation relaxation time. This means that one must
consider two Damk\"ohler numbers in order to quantify the mixing-evaporation dynamics.
We concluded that the simulations of \cite{kumar2018scale} did not exhibit a transition to inhomogeneous mixing with  increasing spatial scale because the supersaturation relaxation time was too small compared to the droplet evaporation time.

Our analysis supports  general conclusions
regarding in-situ observation of droplets in turbulent clouds.
First, most of the local and instantaneous snapshots of droplet configurations observed by \cite{beals2015holographic}
cannot be in the steady states of a global mixing process that mixed undiluted cloud with dry environmental air.
However, a local droplet population may  still be in a local steady state, established as the droplets saturated the air locally.
Such local steady states belong to the transient of a global mixing process. In order to understand the nature of this transient, 
it was necessary  to consider the whole range of possible steady states at different length scales, Figure~\figref{schematic} ({\bf a}).
In short, clouds are not equilibrated at large scales, yet local steady states occur at small scales.

Our analysis also indicates that most of the  droplet populations observed by \cite{beals2015holographic} are likely
to have resulted from mixing with premixed air, and we concluded that the corresponding local steady states  arose by  dilution
rather than complete evaporation. Our model indicates that only very few droplets evaporated completely.

We found that the statistical-model dynamics is somewhat slower than the DNS of \cite{kumar2012extreme,kumar2014lagrangian}, and that the tails of
 our droplet-size distributions are somewhat lighter.
We speculated that this may be due to that the supersaturation dynamics is oversimplified.
Since our model belongs to the family of established PDF models \citep{pope2000turbulent}, it is clear how to address this question in the future \citep{vedula2001dynamics,pope1991mapping,jenny2012modeling}.

Last but not least our analysis highlights which additional observational data is needed for 
 a more quantitative statistical-model analysis of in-situ cloud-droplet measurements. 
 To determine the three key parameters, the volume fraction of cloudy air as well as the two Damk\"ohler numbers,  one needs joint measurements of local droplet populations, supersaturation levels in their vicinity, and the sizes of the local cloud structures. 
  This will allow to characterise and understand the mechanisms underlying local mixing processes observed on different length and time scales, and at different distances from the cloud edge. 
A challenge for the future is to understand the global picture, how evaporation distributes in the cloud, and
where complete droplet evaporation takes place. This is necessary to  improve the \eng{parameterization}{parameterisation} of mixing and evaporation at the cloud edge in sub-grid scale models, in order to  better represent the radiative effects of clouds.
 
\section{Supporting Information}
The SI contains Tables S1-S3 and Appendix S1.
Table S1 lists all parameters of our statistical model simulations. Tables~S2 and S3 specify the DNS results from \cite{kumar2012extreme,Kumar:2013,kumar2014lagrangian,kumar2018scale}, and \cite{Andrejczuk:2006} in Figure~\figref{paramplane}.
Appendix S1 describes the relation between our microscopic Equations~\eqnref{model},
and those used in the DNS of others \citep{vaillancourt2002microscopic,paoli2009turbulent,Lanotte:2009,perrin2015lagrangian,kumar2014lagrangian,kumar2018scale}. We discuss the approximations made in deriving Equations~\eqnref{model}, and quantify their accuracy. We  also describe how to derive the statistical model, Equations \eqnref{statmod}, from Equations~\eqnref{model}, using
the framework of PDF methods \citep{pope2000turbulent}. We detail our computer simulations of the statistical model, and address their numerical convergence. Finally, Appendix S1 contains  details concerning our interpretation of  the  data of \cite{beals2015holographic}, and of direct numerical simulations conducted by other authors.

\section*{ACKNOWLEDGEMENTS}
JF and BM thank B. Kumar and J. Schumacher for providing details of their simulations.
We acknowledge support by Vetenskapsrådet (grant number 2017-368 3865), Formas (grant number 2014-585), and by the grant ‘Bottlenecks for particle growth in turbulent aerosols’ from the Knut and Alice Wallenberg Foundation, Dnr. KAW 2014.0048.
Simulations were performed on resources at Chalmers Centre for Computational Science and Engineering (C3SE) provided by the Swedish National Infrastructure for Computing (SNIC).
\vfill\eject

\section{List of symbols} 
\begin{tabular}{ll}
$t$ & time      \\
$\ve x = (x,y,z)$ & spatial position      \\
$N_0$ & number of droplets at initialisation     \\
$\pe$ & fraction of completely evaporated droplets     \\
$\ve u$ & fluid velocity  \\
$p$ & pressure  \\
$L$ & side length of cubic simulation domain   \\
$w$ & width of initially cloudy region in simulation domain   \\
$s_c$ & supersaturation within initial cloud slab   \\
$s_e$ & supersaturation outside initial cloud slab   \\
$\k$ & turbulent kinetic energy   \\
$U$ & root-mean-square of fluid velocity   \\
$\varepsilon$ & turbulent dissipation rate per unit mass  \\
$\nu$ & kinematic viscosity  \\
$\kappa$ & diffusivity of supersaturation  \\
$s$ & supersaturation  \\
$r$ & droplet radius  \\
$r_0$ & initial volume radius of droplets  \\
$n_0$ & droplet-number density of intially cloudy region  \\
$\overline{r_\alpha (t)s(\ve x_\alpha,t)}$ & average of $r_\alpha (t)s(\ve x_\alpha,t)$ for  droplets in the vicinity of $\ve x$ at time $t$\\
$\tfrac{{\rm D}}{{\rm D}t}$ & Lagrangian time derivative \\
$\tau_{\rm d}$ & droplet evaporation time    \\
$\tau_{\rm s}$ & supersaturation relaxation time    \\
$\tau_\ell$ & time scale for mixing at the length scale $\ell$\\
$\tau_L$ & large-eddy turnover time in simulation domain   \\
${\rm Re}_L$ & turbulence Reynolds number \\
Sc & Schmidt number  \\
$V$ & \dimensionless~volume of simulation domain \\
$\dad$ & Damk\"{o}hler number based on droplet evaporation time  \\
$\das$ & Damk\"{o}hler number based on supersaturation relaxation time  \\
$\ratio$ & Damk\"{o}hler-number ratio  \\
$\ratio_{\rm c}$ & Critical Damk\"{o}hler-number ratio  \\
$\ratio_{\rm min}$ & Lower bound for Damk\"{o}hler-number ratio related to mixing diagrams  \\
$\chi$ & volume fraction of cloudy air     \\
$\chio$ & contribution to the initial volume average of supersaturation      \\
$\theta$ & conserved quantity that reflects the conservation of water and energy  \\
$C_0$, $C_\phi$ & empirical constants   \\
$\left \langle \dots \right \rangle$ & volume average or ensemble average in statistical model   \\
\end{tabular}

\section*{REFERENCES}
\begingroup
\renewcommand{\section}[2]{}

\endgroup

\end{document}


\maketitle
\begin{abstract}
This Supporting Information (SI) contains:
\part*{Appendix S1}
\part*{Tables S1 to S3}
\part*{SI References}
\vspace{1cm}
\end{abstract}

\newpage
\part*{Appendix S1}

\section{Microscopic equations}\label{otherfundamental}
We denote fluid velocity and pressure by $\ve{u}(\ve{x},t)$ and $p(\ve{x},t)$.
Additional fields are the supersaturation $s(\ve{x},t)$, and the condensation rate $\condrate(\ve{x},t)$.
Initially, at $t = 0$,  there are $N_0$ spherical droplets, labeled by $\alpha = 1, \ldots , N_0$,
with  positions $\ve{x}_\alpha(t)$, velocities $\ve{v}_\alpha(t)$, and radii $r_\alpha(t)$.
Our microscopic equations read:
\begin{subequations}
\label{eq:modelsimplified}
\begin{eqnarray}
&&\hspace*{-5mm}\pder{\ve{u}}{t} + (\ve{u}\cdot \nabla) \ve{u}  = - \frac{1}{\airdensity}\nabla p + \nu \nabla^2 \ve{u}, \label{eq:momdimfull} \quad
\nabla\cdot \ve{u} = 0\,, 
\\
&&\hspace*{-5mm}\pder{s}{t} + (\ve{u}\cdot \nabla) s  = \kappa \nabla^2 s - A_2 \condrate(\ve{x},t) \,\label{eq:qvdimfull}
\\
&&\quad \text{with} \quad \condrate(\ve{x},t) = \sum_{\alpha = 1}^{N_0} \kernel (|\ve{x}-\ve{x}_\alpha(t)|) \frac{4\pi}{3} \rhow \der{r_\alpha(t)^3}{t}  \label{eq:gammadimfull} \\
&&\hspace*{-5mm}\frac{{\rm d}\ve{x}_\alpha} {{\rm d}t}= \ve{u}(\ve{x}_\alpha(t),t), \label{eq:droplposdimfull}\\
&&\hspace*{-5mm}\der{r_\alpha^2}{t} = 2 A_3 \,s(\ve{x}_\alpha(t),t). \label{eq:droplraddimfull}
\end{eqnarray}
\end{subequations}
Here, $\airdensity$ is the density of air and $\nu$ is its kinematic viscosity, $\rhow$ is the density of pure liquid water, $\kappa$ is the diffusivity of supersaturation, and $A_2$ and $A_3$ are thermodynamic coefficients.
In Equation~\eqnref{gammadimfull}, $\kernel(|\ve{x}-\ve{x}_\alpha(t)|)$ is a spatial kernel,  \eng{normalized}{normalised} to unity. It makes it possible to construct the continuous condensation-rate field from the dispersed droplets, and ensures that the condensation rate is computed locally \citep{jenny2012modeling}.
The supersaturation is defined as 
\begin{align}\label{eq:supsatdef}
s = e_{\rm v}/e_{\rm vs}(T)-1,
\end{align}
where $e_{\rm v}$ is the partial pressure of water \eng{vapor}{vapour} and $e_{\rm v s}(T)$ is the equilibrium \eng{water-vapor}{water-vapour}  pressure at the temperature $T$ \citep{rogers1989short}. The partial pressure of water \eng{vapor}{vapour} depends on the density $\varrho_{\rm v}$ and gas constant $R_{\rm v}$ of water vapor, and is given by the ideal-gas law \citep{rogers1989short}:
\begin{align}\label{eq:partpressdef}
e_{\rm v} = \varrho_{\rm v} R_{\rm v} T.
\end{align}

Equations~\eqnref{modelsimplified} describe an incompressible flow with advected droplets that condense or evaporate in response to the supersaturation in their vicinity.
Evaporation or condensation causes the droplet radii $r_\alpha$ to decrease or increase, at a rate regulated by the thermodynamic coefficient $A_3$.
The supersaturation $s(\ve x,t)$ evolves according to an advection-diffusion equation, with a source term that describes the response
of  $s(\ve x,t)$ to droplet phase change.
The strength of this response is regulated by the thermodynamic coefficient $A_2$.
The parameters $\airdensity$, $\nu$, $\kappa$, $\rhow$, $A_2$, and $A_3$ are assumed constant in time.
In addition, we impose that a droplet that has evaporated completely must remain at $r_\alpha = 0$.

Equation~\eqnref{droplraddimfull} relies on a scale separation: spatial scales of supersaturation fluctuations induced by turbulent mixing are much larger than the radii of droplets \citep{Vaillancourt:2001}. This means that droplets do not impose boundary conditions on the supersaturation $s(\ve x,t)$ in Equation~\eqnref{qvdimfull}. They do interact locally with the supersaturation field, but over a finite volume determined  by the spatial kernel in Equation~\eqnref{gammadimfull}.
It also means that the supersaturation $s(\ve x,t)$ is not defined on length scales comparable to the radii of droplets, and therefore no a microscopic field in the most strict sense.

The microscopic equations \eqnref{modelsimplified} can be derived from more fundamental descriptions \citep{Vaillancourt:2001,vaillancourt2002microscopic,kumar2014lagrangian,kumar2018scale,perrin2015lagrangian}.
In the following we discuss the assumptions underlying Equations~\eqnref{modelsimplified}.

\subsection{Droplet inertia and settling}
Equation~\eqnref{droplposdimfull}  assumes that the droplets are small enough so that droplet inertia and settling 
are negligible  \citep{vajedi2014clustering,Gus16}.
To neglect droplet inertia is a reasonable approximation when the Stokes number $\text{St} = \tau_{\rm p}/\tau_\eta$
is small enough. Here $\tau_{\rm p} = 2 \rhow r_\alpha^2/(9\airdensity \nu)$ is the Stokes time 
for a small water droplet in air with $\rhow \gg \airdensity$,
and $\tau_\eta = (\nu/\varepsilon)^\frac{1}{2}$ is the Kolmogorov time (of the order of the turnover time of the smallest eddies). 
Settling is negligible at small settling numbers $\text{Sv} = g \tau_{\rm p}/u_\eta$, where $u_\eta = (\nu\varepsilon)^\frac{1}{4}$ is the Kolmogorov velocity and $g > 0$ is the gravitational acceleration.
For typical cloud conditions
 (dissipation rate per unit mass $\varepsilon \sim 10^{-2}$ m$^2$/s$^3$ and viscosity $\nu \sim 1.5\times 10^{-5}$ m$^2$/s),
 the Stokes number is of the order $\text{St} \sim 10^{-4} \, (r_\alpha/\mu\text{m})^2$, and $\text{Sv} \sim 10^{-2} \, (r_\alpha/\mu\text{m})^2$.
In Figure~5 we \eng{analyze}{analyse} observational data from \cite{beals2015holographic}.
Droplet-size distributions in \cite{beals2015holographic} suggest that very few of the measured droplets exceed $r_\alpha = 6\, \mu$m, corresponding to $\text{St} =1.4\times 10^{-2}$
and $\text{Sv} = 2.7 \times 10^{-1}$. Droplet inertia and settling can therefore be neglected for most of these droplets.
Out of the five DNS studies referred to in the main text, \cite{Kumar:2013,kumar2014lagrangian,kumar2018scale} incorporate droplet inertia and settling. Droplet inertia and settling can have minute effects on the droplet-size distribution, despite quite large settling numbers, $\text{Sv} \sim 3.9$ \citep{Kumar:2013}.

\subsection{Buoyancy}
Buoyancy is neglected in the momentum equation, Equation~\eqnref{momdimfull}.
The neglected buoyancy terms read \citep{bannon1996anelastic}:
\begin{align}\label{eq:buoydef}
g B(\ve x,t) = g \Bigg{[} \frac{T-\tempstrat(z)}{\tempstrat(z)} + 0.608 \vapmixrat - \liqmixrat \Bigg{]}.
\end{align}
Here, $B = B(\ve x,t)$ is buoyancy, $T = T(\ve x,t)$ is temperature, and $\tempstrat(z)$ is a static base profile of temperature as a function of altitude $z$.
Furthermore, $\vapmixrat = \vapmixrat(\ve x,t)$ is the \eng{water-vapor}{water-vapour}  mixing ratio, and $\liqmixrat = \liqmixrat(\ve x,t)$ is the liquid-water mixing ratio.
In a dry atmosphere ($\vapmixrat = \liqmixrat = 0$), buoyancy variations are caused only by vertical displacements.
If $\tempstrat(z)$ depends linearly on $z$, then a parcel that is neutrally buoyant at $z = 0$ and is displaced adiabatically to an altitude $z$ has the buoyancy
\begin{align}\label{eq:buoyancy}
B = \frac{1}{\tempstrat(0)} \lp \Gamma - \der{\tempstrat}{z}  \rp z + \mathcal{O}(z^2).
\end{align}
Here, $\Gamma$ is the dry adiabatic lapse rate, the vertical temperature gradient $\mathrm d T/\mathrm d z$ in a dry adiabatic atmosphere at hydrostatic equilibrium \citep{rogers1989short}.
One finds an upper bound for the buoyancy variations within a system of spatial scale $\ell$ by substituting
$z=\ell$ in Equation~\eqnref{buoyancy}. An analogous constraint is derived in Dougherty \citep{dougherty1961anisotropy}. 
Comparing buoyancy acceleration $gB$ at a spatial scale $\ell$ to the inertial acceleration $(\varepsilon^2/\ell)^\frac{1}{3}$ at that scale, one obtains the Dougherty-Ozmidov length scale \citep{grachev2015similarity}:
\begin{align}\label{eq:doughertyscale}
\ell_{\rm B} = \varepsilon^\frac{1}{2}
{\Big(g \frac{1}{\tempstrat(0)}  \left |\Gamma - \der{\tempstrat}{z} \right |  \Big)^{-\frac{3}{4}}}\,.
\end{align}
For typical atmospheric conditions [$\varepsilon \sim 10^{-2}$ m$^2$/s$^3$, $\Gamma \sim -10^{-2}$ K/m, $\mathrm d \tempstrat/\mathrm d z \sim -2\times 10^{-3}$ K/m \citep{dougherty1961anisotropy}], one finds $\ell_{\rm B} \sim$ 45~m.
Under these conditions, buoyancy effects may matter on spatial scales larger than 45 m. 

In a moist atmosphere, buoyancy varies not only as a consequence of vertical displacements, but also because droplets condense and evaporate.
The largest impact of droplet phase change occurs in regions where mixing between cloudy and subsaturated air takes place \citep{vaillancourt2000review}.
In such regions, buoyancy reduces as evaporating droplets absorb latent heat from the air.
\cite{grabowski1993cumulus} show that sedimentation of droplets from 7~cm wide stationary filaments of cloudy air can amplify buoyancy reductions due to droplet evaporation by a factor $\sim 10$ within 5~s. We argue however that such amplifications can not be expected in turbulent clouds, because buoyancy fluctuations at spatial scales $\ell \sim 7$~cm are smoothened by turbulent dissipation at time scales $\tau_\ell \sim (\ell^2/\varepsilon)^{1/3}$ that are smaller than 5~s, already at quite low dissipation rates.
We therefore analyse buoyancy reductions caused by evaporation using the mixing curve of buoyancy \citep{siems1990buoyancy}, which does not take into account droplet sedimentation.
This mixing curve gives an upper bound on the buoyancy reduction $\Delta B$ that phase change induces when a volume fraction $\chi$ of cloudy air is mixed with a volume fraction $1-\chi$ of non-cloudy air.
This upper bound is achieved when the mixture equilibrates at a moist or dry steady state, and it is a definite function of the volume fraction $\chi$, the temperatures $T_\cloudairsubscript$ and $T_\dryairsubscript$, the \eng{water-vapor}{water-vapour}  mixing ratios $\vapmixratcloud$ and $\vapmixratdry$, and the liquid-water mixing ratio $\liqmixratcloud$ \citep{grabowski1993cumulus}.
The subscripts $\cloudairsubscript$ and $\dryairsubscript$ indicate values for the cloudy and non-cloudy mixing substrates.
We estimate $T_{\rm c} \sim 257.9$~K, $\vapmixratcloud \sim 3.1\cdot10^{-3}$, $\liqmixratcloud \sim 5.6\cdot10^{-4}$, $T_{\rm e} \sim 257.6$, and  $\vapmixratdry \sim 2.7\cdot10^{-3}$ for the cloud and cloud environment observed by \cite{beals2015holographic} (see Section~\ref{sec:beals}). From these estimates we compute the upper bound $\Delta B \sim 10^{-3}$ for buoyancy reductions caused by droplet evaporation, using the mixing curve of buoyancy in \cite{grabowski1993cumulus}. Comparing  $g\Delta B$ with the a typical inertial acceleration $(\varepsilon^2/\ell)^\frac{1}{3}$ at the spatial scale $\ell$, we find that buoyancy accelerations due to evaporation are smaller than inertial accelerations for  $\ell<100\,$m.

One caveat is that buoyancy may cause large-scale anisotropies in the fluid motion.
This can introduce additional complexity to droplet evaporation and mixing \citep{perrin2015lagrangian}.
Nevertheless, buoyancy is not a necessary ingredient in a model for these processes.
Buoyancy is neglected in two of the DNS studies that we compare to \citep{kumar2012extreme,Kumar:2013}, as well as in the studies of \cite{pinsky2018theoretical4} and \cite{Jeffery2007}.

\subsection{Temperature changes due to vertical air motion}
Neglecting temperature changes due to vertical air motion is justified for systems that remain at a constant altitude and that are not too large or too moist, as we now explain.
Given the spatial scale $\ell$ of a system that remains at a fixed altitude, the maximal temperature changes caused by vertical air motion are comparable to $ \ell \, \Gamma$.
For the temperature to change by 1 K due to vertical air motion, $\Gamma \sim 10^{-2}$ K/m dictates that we must have $\ell \sim 100$ m.

Temperature regulates droplet evaporation through supersaturation. Supersaturation is determined by the temperature $T$ and the \eng{water-vapor}{water-vapour}  density $\varrho_{\rm v}$ according to Equations~\eqnref{supsatdef} and \eqnref{partpressdef}.
For the values of $\varrho_{\rm v}$ that imply saturation ($s = 0$) at temperatures above 273.15 K, we find that a temperature change of 1 K causes the supersaturation to change by 0.07 or less in magnitude.
In a quite moist system, where the supersaturation is smaller than this in magnitude everywhere, temperature changes due to vertical air motion can be more important than mixing for droplet evaporation.
Since temperature changes due to vertical air motion increase with the spatial scale of the system, results obtained using Equations~\eqnref{modelsimplified} may be inaccurate for systems at constant altitudes that are large and moist.
Temperature changes due to vertical air motion are often negelcted in studies of droplet evaporation \citep{kumar2012extreme,Kumar:2013,kumar2012extreme,kumar2014lagrangian,kumar2018scale,Andrejczuk:2006,pinsky2018theoretical4,siewert2017statistical,Jeffery2007}.

\subsection{Temperature and pressure dependence of the coefficient $A_3$}
We use a constant thermodynamic coefficient $A_3$.
This parameter is given by
\begin{align}\label{eq:athreedef}
A_3 = \Bigg{[} \left ( \frac{\Lambda}{R_{\rm v}T}-1\right ) \frac{\Lambda \rhow}{K_{\rm a} T} + \frac{\rhow R_{\rm v} T}{D_{\rm v} e_{\rm s}(T)}  \Bigg{]}^{-1},
\end{align}
where $\Lambda$ is the latent heat of water vapor, $K_{\rm a}$ is the thermal conductivity of air, and $D_{\rm v}$ is the diffusivity of water \eng{vapor}{vapour} \citep{rogers1989short}.
The parameter $A_3$ decreases with pressure through $D_{\rm v}$, and increases with the temperature $T$ \citep{rogers1989short}.
The values that we use for $A_3$ are computed from temperatures and pressures that are representative of the systems that we consider.
To use a constant value for $A_3$ is usually a good approximation \citep{Lanotte:2009,sardina2015continuous,lehmann2009homogeneous,
kumar2012extreme,andrejczuk2009numerical,siewert2017statistical,
Jeffery2007,pinsky2018theoretical4,devenish2016analytical}.
The reason is that the temperature and pressure dependencies are quite weak.
From Figure~7.1 of \cite{rogers1989short} we see, for example, that $A_3$ increases by 26\% as the temperature increases from 10$^\circ\text{C}$ to 20$^\circ\text{C}$ and the pressure decreases from 100~kPa to 70~kPa.
For observations of a convective cloud at 4000 m altitude \citep{beals2015holographic} we estimate the temperature variations to be smaller than $\sim1^\circ\text{C}$ (see Section~\ref{sec:beals}).
These temperature variations cause $A_3$ to change by less than 10\%, which suggests that it is a good approximation to keep the coefficient constant in analyses of convective clouds at fixed altitude.

\subsection{\eng{Linearization}{Linearisation} of the supersaturation profile}\label{sec:supsatlin}
We treat the joint effect of temperature and water \eng{vapor}{vapour} at the level of a single field, the supersaturation field.
This one-field treatment is obtained by \eng{linearizing}{linearising} the supersaturations dependence on \eng{water-vapor}{water-vapour}  mixing ratio $q_{\rm v}$ and temperature $T$, and by setting the diffusivities of water \eng{vapor}{vapour} and temperature equal to the same value $\kappa$.
With the same diffusivity for temperature and water vapor, and with $s$ given by a linear combination of $\vapmixrat$ and $T$, the two separate advection-diffusion equations for temperature and \eng{water-vapor}{water-vapour}  mixing ratio \citep{Vaillancourt:2001} can be concatenated into a single equation for $s$.

We denote the heat capacity of dry air at constant pressure by $c_{\rm p}$ and estimate $\airdensity \heatcapacity \sim 1000\,$J/(m$^3$K) \citep{rogers1989short}. With this estimate, Table 7.1 of \citep{rogers1989short} implies that the diffusivity of water \eng{vapor}{vapour} roughly equals the diffusivity of temperature
for atmospheric conditions, so that it is sufficient to consider just one value for the diffusivity.
The supersaturation \eng{linearized}{lineqrised} around a reference state with temperature $T_\refsubscript$ and \eng{water-vapor}{water-vapour}  mixing ratio $\vapmixratref$ reads:
\begin{align}\label{eq:subsatlin}
s(\vapmixrat,T) =& s(\vapmixratref,T_\refsubscript) +  \frac{\partial s}{\partial \vapmixrat}  \bigg{|}_{\substack{\vapmixratref\\T_\refsubscript}}(\vapmixrat-\vapmixratref) +  \frac{\partial s}{\partial T}  \bigg{|}_{\substack{\vapmixratref\\T_\refsubscript}}(T-T_{\refsubscript}).
\end{align}
Using $\mathrm d q_{\rm v} = -\mathrm d q_{\rm \ell}$ and $\mathrm d T = (\Lambda/c_{\rm p})\mathrm d q_{\rm \ell}$, we find that the coefficient $A_2$ in Equation~\eqnref{qvdimfull} is given by
\begin{align}\label{eq:a2other}
A_2 = \frac{1}{\airdensity} \Big(\frac{\partial s}{\partial \vapmixrat}  \bigg{|}_{\substack{\vapmixratref\\T_\refsubscript}} - \frac{\Lambda}{\heatcapacity} \frac{\partial s}{\partial T}  \bigg{|}_{\substack{\vapmixratref\\T_\refsubscript}} \Big)\,.
\end{align}
In Figures~2~to~4 we compare our statistical-model results with DNS results \citep{kumar2014lagrangian,kumar2018scale,andrejczuk2009numerical} that are obtained using models where temperature and water \eng{vapor}{vapour} are treated as two separate fields. In order to match their non-linear supersaturation to our linear supersaturation given by Equation~\eqnref{subsatlin}, we \eng{linearize}{linearise} their supersaturation around a saturated reference state $(T_\refsubscript,\vapmixratref)$.
We use $T_\refsubscript = 270.77\,$K \citep{kumar2014lagrangian}, $T_\refsubscript = 282.866\,$K \citep{kumar2018scale} , and $T_\refsubscript = 293\,$K  \citep{andrejczuk2009numerical}, determining our values of $\vapmixratref$ through Equations~\eqnref{supsatdef}, \eqnref{partpressdef}, and $q_{\rm v} = \varrho_{\rm v}/\varrho_{\rm a}$.
For a convective cloud observed by \cite{beals2015holographic} we estimate the temperatures $T_{\rm c} \sim 257.9$ and $T_{\rm e} \sim 257.6$, and the water-vapour mixing ratios $q_{v\rm c} \sim 3.1\times10^{-3}$ and $q_{v\rm e} \sim 2.7\times10^{-3}$, wihtin and outside the cloud (see Section~\ref{sec:beals}).
Over these ranges, the absolue error of the supersaturation linearised at $(T_\refsubscript,\vapmixratref)=(T_{\rm c},q_{v\rm c})$ is less than 1\%. The relative error at $(T_{\rm e},q_{v\rm e})$ is around 2\%.

\section{Initial conditions}\label{sec:cloudconfig}
To allow for quantitative comparisons with the results of \cite{Kumar:2013,kumar2012extreme,kumar2014lagrangian,kumar2018scale}, we use the same (or almost identical) initial conditions:
 a slab of cloudy air inside a cubic domain of size $L$ with periodic boundary conditions. The cloudy air occupies a fraction $\chi$ of the domain, and it is surrounded by subsaturated air [Figure~1({\bf b})]. The initial droplet-size distribution is either monodisperse or Gaussian, with mean $r_\mu$ and standard deviation $\sigma_0$.
When we compare to DNS from \cite{Kumar:2013,kumar2012extreme,kumar2014lagrangian,kumar2018scale} in Figures~2~to~4, we use their initial droplet-size distribution, which is monodisperse ($\sigma_0 = 0$). When we \eng{analyze}{analyse}  measurements of \cite{beals2015holographic}, we use the Gaussian droplet-size distribution of the undiluted cloud that corresponds to the measurements ($\sigma_0 \neq 0$, see Section \ref{sec:params}).
The simulations of \cite{Andrejczuk:2006} are for decaying turbulence with quite strong buoyancy effects, and a
different  initial geometry.
Nevertheless, some of our results can be compared qualitatively to the results of \cite{Andrejczuk:2006}, as explained in Section \ref{sec:DNS}.
Our analysis of the measurements of \cite{beals2015holographic} is based on our initial condition.
This analysis rests upon the assumption that our system models a structure of size $L$ formed at the  cloud edge,
although  its detailed composition and geometry can deviate from our slab-like initial conditions. 

Our initial conditions are homogeneous in two of the three coordinate axes.
We take $x$ to denote the coordinate in the spatially inhomogeneous direction.
Initially the droplets are distributed uniformly and randomly within a three-dimensional cloud slab, the volume for which $-\chi L/2 < x < \chi L/2$ [Figure~1({\bf b})].
The initial droplet-number density of the slab is  $n_0 = N_0/(\chi V)$, where $V = L^3$ is the domain volume.
To obtain the statistical-model results shown in Figures~2~to~4 we use the initial supersaturation profiles of \cite{Kumar:2013,kumar2012extreme,kumar2014lagrangian,kumar2018scale}:
\begin{align}\label{eq:hs}
s(x,0) = (s_\cloudairsubscript-s_\dryairsubscript) \exp \Bigg{[} -\profilefactor  \lp \frac{x}{L}\rp^{\profileexponent} \Bigg{]}\, + s_\dryairsubscript.
\end{align}
Here, $s_\cloudairsubscript > 0$ and $s_\dryairsubscript < 0$ denote the initial supersaturation at the center of the slab and of the dry air [Figure~1({\bf b})].
The shape of the initial supersaturation profile is contained in the parameters $\profilefactor$ and $\profileexponent$. 
The homogeneous mixing line of \cite{beals2015holographic} passes through the top-right corner of the mixing diagram in Figure~5({\bf a}).
To be able to compare with this homogeneous mixing line, we use a sharp  initial supersaturation profile with $s_\cloudairsubscript =0$ when \eng{analyzing}{analysing} the measured data:
\begin{align}\label{eq:hssimple}
s(x,0) = \begin{cases}
0 \quad \,\,\text{for} \quad |x/L| < \chi/2\,, \\
s_\dryairsubscript \quad \text{otherwise}\,. \quad
\end{cases}
\end{align}

\section{\Dimensionless~equations and parameters}\label{sec:dimensionless}
We \dedimensionalise~ Equations~\eqnref{modelsimplified} using $U = \sqrt{2\,\k/3}$ and the large-eddy time $\tau_L = \k/\varepsilon$, which is proportional to $L/U$ if the size of the largest eddies scale with $L$ \citep{pope2000turbulent}.
In addition, we use the following quantities to \dedimensionalise: the droplet-number density of the slab [$n_0 = N_0/(\chi V)$], the (positive) supersaturation $|s_\dryairsubscript|$, and the initial volume radius
\begin{align}
r_0 = \left [ \frac{1}{N_0}\sum_{\alpha = 1}^{N_0} r_\alpha(0)^3 \right ]^\frac{1}{3}.
\end{align}
We \dedimensionalise~as follows: $\ve{u}' = \ve{u}/U$, $\ve{x}' = \ve{x}/(U\tau_L)$, $t' = t/\tau_L$, $p' = p/(\airdensity U^2)$, $\ve{x}_\alpha' = \ve{x}_\alpha/(U\tau_L)$, $s' = s/|s_\dryairsubscript|$, $r_\alpha' = r_\alpha/r_0$, $\kernel' = \kernel (U\tau_L)^3$, $n' = n/n_0$, $s_\cloudairsubscript' = s_\cloudairsubscript/|s_\dryairsubscript|$, $\sigma_0' = \sigma_0/r_0$, $L' = L/(U\tau_L)$ and $V' = V/(U\tau_L)^3$.
Dropping the primes,  Equations~\eqnref{modelsimplified} take the \dimensionless~form:
\begin{subequations}
\label{eq:modelsimplifieddl}
\begin{eqnarray}
&&\hspace*{-5mm}\pder{\ve{u}}{t} + (\ve{u}\cdot \nabla) \ve{u}  = - \nabla p + \frac{1}{\turbreynolds} \nabla^2 \ve{u}\,, \label{eq:veldl} \\
&&\hspace*{-5mm}\nabla\cdot \ve{u} = 0\,, \label{eq:incomprdl} \\
&&\hspace*{-5mm}\pder{s}{t} + (\ve{u}\cdot \nabla) s  = \frac{1}{\turbreynolds {\rm Sc}} \nabla^2 s - \das \chi V \overline{r_\alpha(t) s(\ve{x}_\alpha,t)}\,, \label{eq:sdl}\\
&&\hspace*{-5mm}\frac{{\rm d}\ve{x}_\alpha} {{\rm d}t}= \ve{u}(\ve{x}_\alpha(t),t)\,, \label{eq:droplposdl}\\
&&\hspace*{-5mm}\der{r_\alpha^2}{t} = \dad \,s(\ve{x}_\alpha(t),t) \label{eq:droplradl}\,.
\end{eqnarray}
\end{subequations}
Here $\overline{r_\alpha(t) s(\ve{x}_\alpha,t)}$ denotes the average of $\kernel(|\ve{x}-\ve{x}_\alpha(t)|) r_\alpha(t) s(\ve{x}_\alpha,t)$ over the droplets:
\begin{align}\label{eq:corrdl}
\overline{r_\alpha(t) s(\ve{x}_\alpha,t)} = \frac{1}{N_0} \sum_{\alpha=1}^{N_0} \kernel(|\ve{x}-\ve{x}_\alpha(t)|) r_\alpha(t) s(\ve{x}_\alpha,t).
\end{align}
The average $\overline{r_\alpha(t) s(\ve{x}_\alpha,t)}$ is position dependent, it depends on number
and sizes of the local droplets, and upon the local supersaturation $s(\ve{x}_\alpha,t)$ tied to these droplets. The dynamics in Equations~\eqnref{modelsimplifieddl} is identical to the dynamics in Equations~\eqnref{modelsimplified}, but \dimensionless.

Equations~\eqnref{modelsimplifieddl} have the following \dimensionless~parameters: the domain volume $V$, the initial cloud fraction $\chi$, the turbulence Reynolds number $\turbreynolds = \tfrac{2}{3}\k^2/(\varepsilon \nu)$ \citep{pope2000turbulent}, the Schmidt number $\text{Sc} = \nu/\kappa$ [of order unity since the diffusivities of
both temperature and water \eng{vapor}{vapour} are roughly the same as the kinematic viscosity of air \citep{perrin2015lagrangian}], and the two Damk\"{o}hler numbers
\begin{align}\label{eq:das}
\das = \tau_L/\tau_\text{s} \quad \text{and}  \quad \dad = \tau_L/\tau_\text{d},
\end{align}
where $\tau_\text{s}$ and $\tau_\text{d}$ denote the supersaturation relaxation time and the droplet evaporation time.
These time scales are given by
\begin{align}\label{eq:taus}
\tau_\text{s} = (4 \pi A_2 A_3 \rho_\text{w} n_0 r_0)^{-1} \quad \text{and} \quad \tau_\text{d} = \frac{r_0^2}{2 A_3 |s_\dryairsubscript|}.
\end{align}

The time scales $\tau_\text{s}$ and $\tau_\text{d}$ can be very different. This can be understood by \eng{analyzing}{analysing} their ratio 
\begin{align}\label{eq:ratiorhol}
\ratio = \frac{\tau_\text{s}}{\tau_\text{d}} = \frac{\dad}{\das}=\frac{|s_\dryairsubscript|}{2 \pi A_2 \rho_\text{w} n_0 r_0^3} = \frac{2 |s_\dryairsubscript|}{3 A_2 \varrho_{\ell 0}}.
\end{align}
Here, we introduce $\varrho_{\ell 0} = 4 \pi r_0^3 n_0 \rhow/3$ as a scale for liquid water density (mass per volume), the liquid water density in a cloud with droplet-number density $n_0$ and mean volume radius $r_0$. The scale for liquid water density determines $\ratio$ together with the supersaturation $s_\dryairsubscript$ of the dry mixing substrate, and the thermodynamic coefficient $A_2$.
For typical cloud conditions, we find $A_2 \sim 260\,$m$^3$/kg using Equation~\eqnref{a2other}.

Consider, for example, a cumulus cloud with a typical \citep{rogers1989short} liquid water density $0.3$~g/m$^3$ in the bulk that contains premixed and slightly subsaturated air with supersaturation $-0.01$.
Substituting $s_\dryairsubscript = -0.01$ and $\varrho_{\ell0} = 0.3$~g/m$^3$ into Equation~\eqnref{ratiorhol}, we find that $\ratio \sim 0.09$ for the local mixing processes in the bulk, so  $\tau_\text{d}$ is more than ten times larger than $\tau_\text{s}$.
At the cloud edge, by contrast, the liquid water density is lower, and mixing with dry air occurs. Assuming, say, $s_\dryairsubscript = -0.1$ and $\varrho_{\ell0} = 0.1$~g/m$^3$ for a local mixing process yields $\ratio = 2.6$, so $\tau_\text{d}$ is smaller than $\tau_\text{s}$. As we continue to move away from the cloud core, $\varrho_{\ell0}$ tends to zero, and $\tau_\text{s}/\tau_\text{d}$ tends to infinity. In conclusion, the supersaturation relaxation time and the droplet evaporation time for local mixing processes can differ by orders of magnitude.\\[0.1cm]

\section{Statistical model}
The statistical model, Equations~\maintextstatmodeqnnum, rests upon on a probabilistic description of the microscopic Equations~1, in terms of two probability-density functions (PDF:s). The first PDF, denoted by $\mathscr{F}$, describes the droplets, and the second one, denoted by $f$, describes the air.
The dynamics of Lagrangian fluid elements in Equations~\maintextstatmodeqnnum~constitutes a closed set of evolution equations for these PDF:s.
The presence of droplets makes it necessary to consider two types of fluid elements, one for each PDF.
Similar extensions of single-phase PDF models have already been made to describe combustion problems, where a gas interacts with droplets or other dispersed particles \citep{jenny2012modeling,stollinger2013pdf}.

In the following, we derive the statistical model in terms of $\mathscr{F}$ and $f$, since this allows us to highlight similarities and differences to PDF models for single-phase flows \citep{pope1985pdf}.
The first step is to define the two PDF:s.
Their exact evolution equations are not closed, as they contain conditional averages that are not known in terms of the PDF:s.
We employ standard closures, designed to approximate the effects of acceleration and diffusive scalar flux in single-phase flows \citep{pope1985pdf} and multiphase combustion \citep{jenny2012modeling,stollinger2013pdf,haworth2010progress}.
The resulting closed set of equations constitutes a statistical model that describes the dynamics dictated by the microscopic Equations~1.
As a final step, the model is cast into the form of Langevin equations \citep{pope2000turbulent}, invoking the concept of Lagrangian fluid elements.

\subsection{Probabilistic description}
We describe the droplets by the joint PDF of droplets and air, $\mathscr{F}(\ve U,S,R^2,\ve{x};t)$, which is the probability density of the event
\begin{align}\label{eq:lagrdef}
E_{\rm d}: \{ \ve x_\alpha(t) = \ve{x}, \quad  r_\alpha(t)^2 = R^2 > 0, \quad \ve{u}(\ve{x}_\alpha(t),t) =  \ve{U}, \quad \text{and} \quad s(\ve{x}_\alpha(t),t) =  S\, \}
\end{align}
for a droplet $\alpha$. Our definition of probability densities of events is that of \cite{pope1985pdf}.
The PDF $\mathscr{F}$ is {\em Lagrangian}, since it is a PDF evaluated along Lagrangian trajectories \citep{pope2000turbulent}, the trajectories of the advected droplets.
By defining $\mathscr{F}$ as a density in squared droplet radius $r_\alpha^2(t)$, instead of droplet radius $r_\alpha(t)$, we avoid complications that follow from that $\mathrm d r_\alpha/\mathrm d t$ diverges in the limit $r_\alpha(t) \rightarrow 0$ when we formulate the evolution equation of $\mathscr{F}$ below.
Droplets that evaporate according to Equation~\eqnref{droplradl} may adopt a zero radius, $r_\alpha^2(t)=0$.
When $r_\alpha^2(t)$ becomes zero the droplet has evaporated completely and no longer plays any role for the dynamics.
Therefore only droplets with $R^2>0$ are considered in $\mathscr{F}$. In this setup $\mathscr{F}$ is  \eng{normalized}{normalised} to $1 - P_{\rm e}(t)$, reflecting that the total probability of $\mathscr{F}$ decreases as droplets completely evaporate.
The observables that we solve for are contained in $\mathscr{F}$. We compute the droplet-size distribution as
\begin{align}\label{eq:dsdstatmod}
\mathscr{F}_r(R;t) = 2 R \int \mathscr{F}(\ve U,S,R^2,\ve{x};t) \mathrm d \ve U \; \mathrm d S \; \mathrm d \ve x,
\end{align}
and we compute the fraction of completely evaporated droplets as
\begin{align}\label{eq:pestatmod}
P_\esubscript(t) = 1 - \int \mathscr{F}(\ve U,S,R^2,\ve{x};t) \mathrm d \ve U \mathrm d S \mathrm d R^2 \mathrm d \ve x.
\end{align}

The air is described by the joint PDF of velocity and supersaturation, $f(\ve U,S;\ve x,t)$, which is the probability density of the event
\begin{align}\label{eq:eulerdef}
E_{\rm a}: \{\ve u(\ve x,t) = \ve{U} \quad \text{and} \quad s(\ve x,t) = S\, \}.
\end{align}
As opposed to $\mathscr{F}$, the PDF $f$ is  \eng{normalized}{normalised} to unity at all times. Note also that $f$ is {\em Eulerian}, since it is a PDF of fluid properties at a fixed position \citep{pope2000turbulent}.
PDF models for single-phase flows describe turbulent flows with one or several advected scalar fields \citep{pope2000turbulent}.
They solve for an Eulerian joint PDF of velocity and scalars. This PDF is analogous to $f$, but may describe more than one advected scalar field.
As in some combustion problems \citep{stollinger2013pdf,jenny2012modeling,haworth2010progress}, we must consider a dispersed phase (the droplets) that interacts with a fluid phase (the air). As a consequence, we need two PDF:s to describe the dynamics.

Exact evolution equations of Lagrangian and Eulerian PDF:s are given in \cite{pope1985pdf}.
To obtain the evolution equation for $\mathscr{F}$, we integrate a corresponding transition PDF in \cite{pope1985pdf} over $\mathscr{F}(\ve U,S,R^2,\ve{x};t=0)$, which is known from the intial conditions.
The microscopic equations \eqnref{modelsimplifieddl} imply that the dynamics of $\mathscr{F}$ and $f$ read:
\begin{adjustwidth}{-70pt}{-10pt}
\begin{subequations}
\label{eq:pdfmodel}
\begin{eqnarray}
&&\pder{\mathscr{F}}{t} + U_i\pder{\mathscr{F}}{x_i}+   \dad  S  \pder{\mathscr{F}}{R^2}  \notag \\ 
&&\hspace{10mm}=  - \pder{}{U_i} \la \frac{1}{\turbreynolds } \nabla^2 u_i -\pder{p}{x_i}\vline \hspace{1pt}E_{\rm d} \ra \mathscr{F}- \pder{}{S} \la \frac{1}{\turbreynolds {\rm Sc}} \nabla^2 s\vline \hspace{1pt}E_{\rm d} \ra \mathscr{F} - \das \chi V \pder{}{S} \la \overline{r_\alpha(t)s(\ve x_\alpha,t)} \vline\hspace{1pt}E_{\rm d} \ra \mathscr{F}\,
 \label{eq:droplpdfuncl}, \\
 \notag \\
&&\pder{f}{t} + U_i\pder{f}{x_i}  = - \pder{}{U_i} \la \frac{1}{\turbreynolds } \nabla^2 u_i -\pder{p}{x_i}\vline \hspace{1pt}E_{\rm a} \ra f- \pder{}{S} \la \frac{1}{\turbreynolds {\rm Sc}} \nabla^2 s\vline \hspace{1pt}E_{\rm a} \ra f - \das \chi V \pder{}{S} \la \overline{r_\alpha(t)s(\ve x_\alpha,t)} \vline\hspace{1pt}E_{\rm a} \ra f\hspace{1pt}\, \label{eq:airpdfuncl}.
\end{eqnarray}
\end{subequations}
\end{adjustwidth}
Here, we follow \cite{pope1985pdf} and write the components of positions and velocities by $(x_1,x_2,x_3)$, $(u_1,u_2,u_3)$ and $(U_1,U_2,U_3)$, and sum over repeated indices.
Averages conditioned upon the events in Equations~\eqnref{lagrdef} and \eqnref{eulerdef} are denoted by $\la \ldots | E_{\rm d} \ra$ and $\la \ldots | E_{\rm a} \ra$.

We have imposed that a droplet that has evaporated completely must remain at $r_\alpha^2(t) = 0$.
Since $\mathscr{F}$ describes  droplets with positive radii, this dynamics implies a boundary condition for $\mathscr{F}$ at $R^2 = 0$, described by the probability current \citep{gardiner2009stochastic}
\begin{align}\label{eq:currdef}
\mathscr{J}_{\rm e}(\ve U,S,\ve{x};t) =\begin{cases}
- \dad S  \Lim{R^2\rightarrow 0^+}\mathscr{F}(\ve U,S,R^2,\ve{x};t) \quad&\text{if}\quad{S < 0}\\
0 & \text{otherwise}
\end{cases}
\end{align}
in the negative $R^2$ direction.
Here, the limit $R^2\rightarrow 0^+$ is required, since $R^2 > 0$ in $\mathscr{F}$.
The probability current in Equation~\eqnref{currdef} is non-negative, $\mathscr{J}_{\rm e}(\ve U,S,\ve{x};t) \geq 0$, which reflects that droplets can reach $r_\alpha^2(t) = 0$ from above, but not from below.
The rate of complete droplet evaporation can be written in terms of the probability current as
\begin{align}\label{eq:decomp}
\der{P_{\rm e}}{t} = \int \mathscr{J}_{\rm e}(\ve U,S,\ve{x};t) \mathrm d \ve U\mathrm d S \mathrm d\ve{x}.
\end{align}
In addition to the boundary condition at $R^2 = 0$, $\mathscr{F}$ and $f$ inherit the spatial periodicity of the slab configuration [Figure~(1{\bf b})].
The variables $\ve U$ and $S$ are not bounded, and require no boundary conditions.
The initial conditions for $\mathscr{F}$ and $f$ are determined by the initial supersaturation profile (Equation~\eqnref{hs}~or~\eqnref{hssimple}), the initial spatial distribution of droplets, the initial droplet-size distribution, and the PDF of velocity $\varphi(\ve U;\ve x,t)$, which is the probability density of the event $\ve u(\ve x,t) = \ve U$.
We discuss the PDF $\varphi(\ve U;\ve x,t)$ below.

The evolution equations \eqnref{pdfmodel}, together with their initial and boundary conditions, constitute an exact description of the dynamics of $\mathscr{F}$ and $f$.
The terms on the left-hand sides of Equations~\eqnref{droplpdfuncl} and \eqnref{airpdfuncl} are in closed form, as they can be computed in terms of the PDF:s.
The terms on the right-hand sides are not in closed form, and must be approximated in order to solve for the joint evolution of $\mathscr{F}$ and $f$.
They correspond to the fluctuating acceleration and diffusive flux of supersaturation that a fluid element experiences.

\subsection{Closure}
We follow \citep{pope2000turbulent} and model the fluctuating acceleration of fluid elements with velocities $\ve {u}(t)$ in statistically stationary, homogeneous and isotropic turbulence as a three-dimensional Ornstein-Uhlenbeck process:
\begin{align}
\mathrm d \ve u = -\frac{3}{4}C_0 \ve u \mathrm d t  + \left ( \frac{3}{2} C_0 \right )^\frac{1}{2} \mathrm d  \ve \noise \,. \label{eq:velmcsuppl}
\end{align}
In this Langevin equation, $C_0$ is a constant that depends on the Taylor-scale Reynolds number $\text{Re}_\lambda =\sqrt{10\turbreynolds }$ \citep{pope2011simple}, and $\mathrm d \ve \eta$ is the increment of a three-dimensional isotropic Wiener process \citep{pope2000turbulent}.
We model the $\text{Re}_\lambda$-dependence of $C_0$ according to the empirical fit of  \cite{pope2011simple}, $C_0 = 6.5/(1+140 \text{Re}_\lambda^{-4/3})^\frac{3}{4}$.
Equation~\eqnref{velmcsuppl} dictates that each velocity component of a fluid element evolves stochastically according to an independent one-dimensional Ornstein-Uhlenbeck process with \dimensionfull~variance $2\k/3$.
Another consequence of Equation~\eqnref{velmcsuppl} is that the auto-correlation function $\rho(s) = \la u(t)u(t+s)\ra$ (\dimensionfull~$t$ and $s$) of the velocity component $u(t)$ of a fluid element decays exponentially at the time-scale $4 \tau_L/(3 C_0)$.
In our \dimensionless~units, the Ornstein-Uhlenbeck processes have variance unity and auto-correlation time $4 /(3 C_0)$.
With Equation~\eqnref{velmcsuppl}, the PDF of fluid-element velocity relaxes to a joint normal distribution at this time scale \citep{pope2000turbulent}.

The joint normal PDF of fluid-element velocity components and the exponential decay of $\rho(s)$ dictated by Equation~\eqnref{velmcsuppl} is observed empirically \citep{pope2000turbulent,pope2011simple}. In the limit $\text{Re}_\lambda \rightarrow \infty$, $C_0$ approaches 6.5, a measured value for the Kolmogorov constant that specifies how the second order Lagrangian structure function depends on the Lagrangian auto-correlation time and the mean dissipation rate in high-Reynolds number turbulence \citep{pope2011simple}. Consistently with the Kolmogorov theory of turbulence \citep{kolmogorov1941local}, the statistics pertaining to the motion of a fluid element subject to Equation~\eqnref{velmcsuppl} is Reynolds-number independent at high Reynolds numbers.
Equation~\eqnref{velmcsuppl} does not take into account that the dissipation rate experienced by a fluid element fluctuates intermittently around the mean dissipation rate $\varepsilon$, as predicted by Kolmogorov's refined theory of turbulence \citep{kolmogorov1962refinement}. This intermittency can be taken into account using the refined Langevin model of \cite{pope1990velocity}, which is an extension of Equation~\eqnref{velmcsuppl}.
Since Equation~\eqnref{velmcsuppl} reproduces the observed variance and autocorrelation time of fluid-element velocities,  Equation~\eqnref{velmcsuppl}  ensures that the empirically observed turbulent diffusion of passive scalars is correctly described \citep{pope2000turbulent}.

Equation~\eqnref{velmcsuppl} for the velocity of fluid elements provides closure for the first terms on the right-hand sides of Equations~\eqnref{droplpdfuncl} and \eqnref{airpdfuncl}.
The joint PDF of droplets and air $\mathscr{F}$ is a probability density in the velocity of a fluid element, a fluid element that moves together with one of the advected droplets. The closure for Equation~\eqnref{droplpdfuncl} therefore follows directly from Equation~\eqnref{velmcsuppl}.
The joint PDF of velocity and supersaturation $f$ is not a probability density in the velocity of a fluid element. The closure implied for Equation~\eqnref{airpdfuncl} is therefore somewhat more intricate.
As explained by \cite{pope2000turbulent}, the crucial step in the derivation of this closure is to consider the relation between (the Eulerian) $f$ and a corresponding Lagrangian PDF that is conditioned upon the initial position of a fluid element.
In an incompressible flow, $f$ is obtained by integrating this Lagrangian PDF over initial positions of fluid elements.
Equation~\eqnref{velmcsuppl} prescribes a Fokker-Planck equation for the Lagrangian PDF. 
Since this Fokker-Planck equation does not depend upon the initial position of the fluid element, a Fokker-Planck equation of the same form follows for $f$.
Following \cite{pope2000turbulent}, we conclude that Equation~\eqnref{velmcsuppl} implies that the first terms on the right-hand sides of Equations~\eqnref{droplpdfuncl} and \eqnref{airpdfuncl} are given by two Fokker-Planck equations:
\begin{subequations}
\label{eq:velpdfclosure}
\begin{eqnarray}
&&- \pder{}{U_i} \la \frac{1}{\turbreynolds } \nabla^2 u_i -\pder{p}{x_i}\vline \hspace{1pt}E_{\rm d} \ra \mathscr{F}=\frac{3}{4} C_0  \pder{}{U_i}    U_i \mathscr{F} + \frac{3}{4} C_0 \pder{^2\mathscr{F}}{U_i \partial U_i}, \quad \text{and} \\
 \notag \\
&&- \pder{}{U_i} \la \frac{1}{\turbreynolds } \nabla^2 u_i -\pder{p}{x_i}\vline \hspace{1pt}E_{\rm a} \ra f=\frac{3}{4} C_0  \pder{}{U_i}    U_i f + \frac{3}{4} C_0 \pder{^2f}{U_i \partial U_i}\, \label{eq:fdsfsdf}.
\end{eqnarray}
\end{subequations}
In addition to providing closure to the first terms on the right-hand sides of Equations~\eqnref{droplpdfuncl} and \eqnref{airpdfuncl}, Equation~\eqnref{velmcsuppl} sets the initial conditions of $\mathscr{F}$ and $f$. Statistical homogeneity and stationarity dictates that the PDF of velocity $\varphi (\ve U;\ve x,t)$ discussed above is independent of $\ve x$ and $t$. To obtain statistically stationary modeled turbulence it follows from Equation~\eqnref{velmcsuppl} that $\varphi (\ve U;\ve x,t)$ must be joint normal in the velocity components.

For the second terms on the right-hand sides of Equations~\eqnref{droplpdfuncl} and \eqnref{airpdfuncl}, we make the following closure:
\begin{align}
&&\la \frac{1}{\turbreynolds {\rm Sc}} \nabla^2 s\vline \hspace{1pt}E_{\rm d} \ra = \la \frac{1}{\turbreynolds {\rm Sc}} \nabla^2 s\vline \hspace{1pt}E_{\rm a} \ra = -\frac{1}{2} C_\phi (S-\la s(\ve x,t)\ra). \label{eq:diffclosure}
\end{align}
Here, $C_\phi$ is an empirical constant, and $\la s(\ve x,t) \ra$ is defined as the position-dependent average
\begin{align}\label{eq:smeandef}
\la s (\ve x,t)\ra = \int S f(\ve U,S;\ve x,t) \mathrm d \ve U\, \mathrm d S.
\end{align}
Equation~\eqnref{diffclosure} ensures that the combined effect of molecular diffusion and turbulent mixing results in the correct decay of the variance ${\rm Var} [ s(\ve x,t) ] = \la s (\ve x,t)^2\ra - \la s (\ve x,t)\ra^2$ of supersaturation, at least under homogeneous conditions ($\partial f/\partial x_i = 0$). Here, $\la s (\ve x,t)^2\ra$ is the mean-squared supersaturation, given by Equation~\eqnref{smeandef} with $S$ replaced by $S^2$ in the integrand. Experiments and DNS support that, under homogeneous conditions, the variance of supersaturation in fully developed turbulence decays as
\begin{align}\label{eq:supsatvardecay}
\frac{\mathrm d}{\mathrm d t} {\rm Var}  [ s(\ve x,t) ] =- C_\phi {\rm Var} [ s(\ve x,t) ],
\end{align}
with $C_\phi \sim 2$, independently of the molecular diffusivity (independently of $\rm Sc$) \citep{pope2000turbulent}. This decay rate is reproduced by Equation~\eqnref{diffclosure}.
Furthermore, Equation~\eqnref{diffclosure} respects the boundedness condition of a passive scalar \citep{pope2000turbulent}. This means that, in the absence of evaporation or condensation, the supersaturation remains bounded between its maximal and minimal values.
Other aspects of the decay of supersaturation fluctuations may not be correctly reproduced by Equation~\eqnref{diffclosure} \citep{pope2000turbulent}.

For the third terms on the right-hand sides of Equations~\eqnref{droplpdfuncl} and \eqnref{airpdfuncl}, we make the following closure:
\begin{align}
\la \overline{r_\alpha(t)s(\ve x_\alpha,t)} \vline\hspace{1pt}E_{\rm d} \ra = \la \overline{r_\alpha(t)s(\ve x_\alpha,t)} \vline\hspace{1pt}E_{\rm a} \ra =  \la r(t) s(\ve x,t)\ra\, \label{eq:evapclosure},
\end{align}
where $\la r(t) s(\ve x,t) \ra$ is the position-dependent average
\begin{align}
\la r(t) s(\ve x,t) \ra = \int R S \mathscr{F}(\ve U,S,R^2,\ve{x};t) \mathrm d \ve U\, \mathrm d S \,\mathrm d R^2.
\end{align}
Equation~\eqnref{evapclosure} ensures that the liquid-water potential temperature analogue $\theta$ is conserved under the statistical-model. 
With this closure we replace the local average $\overline{r_\alpha(t) s(\ve{x}_\alpha,t)}$, which describes local neighborhoods of droplets, by $\la r(t) s(\ve{x},t) \ra$, which is an expectation for a single droplet.
As a consequence, the self-limiting nature of droplet evaporation may not be accurately described by the statistical model.
To model local conditional averages as in Equation~\eqnref{evapclosure} is a standard way to obtain closure for PDF models that describe combustion of particles in turbulence \citep{jenny2012modeling,stollinger2013pdf,haworth2010progress}.

Inserting Equations~\eqnref{velpdfclosure}, \eqnref{diffclosure} and \eqnref{evapclosure} into Equations~\eqnref{pdfmodel}, we obtain a model for the joint evolution of $\mathscr{F}$ and $f$:
\begin{adjustwidth}{-70pt}{-10pt}
\begin{subequations}
\label{eq:pdfmodelclosed}
\begin{eqnarray}
&&\pder{\mathscr{F}}{t} + U_i\pder{\mathscr{F}}{x_i}  + 2 \dad S \pder{\mathscr{F}}{R^2}  \notag \\ 
&&\hspace{10mm}=  \frac{3}{4} C_0  \pder{}{U_i}    U_i \mathscr{F} + \frac{3}{4} C_0 \pder{^2\mathscr{F}}{U_i \partial U_i} +  \frac{1}{2} C_\phi \pder{}{S} (S-\la s(\ve x,t)\ra) \mathscr{F} - \das \chi V \pder{}{S} \la r(t) s(\ve x,t)\ra \mathscr{F}\,
 \label{eq:droplpdfunclclosed},  \\
 \notag \\
&&\pder{f}{t} + U_i\pder{f}{x_i}  =  \frac{3}{4} C_0  \pder{}{U_i}    U_i f + \frac{3}{4} C_0 \pder{^2f}{U_i \partial U_i} +  \frac{1}{2} C_\phi \pder{}{S} (S-\la s(\ve x,t)\ra) f - \das \chi V \pder{}{S} \la r(t) s(\ve x,t)\ra f \, \label{eq:airpdfunclclosed}.
\end{eqnarray}
\end{subequations}
\end{adjustwidth}
These equations are the statistical model, described as a closed set of evolution equations for $\mathscr{F}$ and $f$.
We note that $\mathscr{F}$ and $f$ couple to each other through the averages $\la s(\ve x,t) \ra$ and $\la r(t)s(\ve x,t) \ra$.
This coupling is a consequence of that droplets are affected by the supersaturation of the air, and that the supersaturation of the air is affected by evaporating droplets.

\subsection{Lagrangian dynamics}
The PDF dynamics in Equations~\eqnref{pdfmodelclosed} can be cast into a set of equations that describe the dynamics of Lagrangian fluid elements \citep{pope2011simple}.
The dynamics of Lagrangian fluid elements with positions $\ve x(t)$, velocities $\ve u(t)$, and supersaturations $s(t)$ that corresponds to Equation~\eqnref{airpdfunclclosed} is given by Equation~\eqnref{velmcsuppl}, together with:
\begin{subequations}
\label{eq:statmodsuppl}
\begin{eqnarray}
&&\hspace*{-5mm}\der{\ve x}{t}   = \ve u \label{eq:posmcsuppl}\, ,\\
&&\hspace*{-5mm}\der{s}{t}   = - \frac{1}{2}C_\phi \left  ( s -\langle s(\ve x,t)\rangle \right ) - \das\, \chi V \langle r(t) s( \ve x,t)\rangle\,  \label{eq:supsatmcsuppl}.
\end{eqnarray}
\end{subequations}
Since droplets are advected, they move together with Lagrangian fluid elements. Therefore, some Lagrangian fluid elements coincide with droplets.
We describe such fluid elements by the radii $r(t)$ of the droplets that they coincide with, in addition to their positions, velocities and supersaturations.
Their dynamics corresponds to Equation~\eqnref{droplpdfunclclosed}, and is given by Equations~\eqnref{velmcsuppl} and \eqnref{statmodsuppl}, together with
\begin{align}
\der{r^2}{t}   = \begin{cases}
\dad s \quad \text{if} \quad r^2(t) > 0\\
0 \quad \hspace{6mm}\text{if} \quad r^2(t) = 0
\end{cases} \label{eq:radmcsuppl}.
\end{align}

The relations between fluid-element dynamics and evolution equations of Eulerian and Lagrangian PDF:s are derived by \cite{pope1985pdf}.
To conclude why Equations~\eqnref{velmcsuppl}, \eqnref{statmodsuppl}, and \eqnref{radmcsuppl} correspond to Equations~\eqnref{pdfmodelclosed}, we give a brief summary of the derivations for our statistical-model dynamics.
First, one \eng{recognizes}{recognises} that fluid elements with droplets sample $\mathscr{F}$, and that fluid elements without droplets sample $f$. Second, one formulates the Fokker-Planck equations for the two types of fluid elements.
The Fokker-Planck equation for fluid elements with droplets is Equation~\eqnref{droplpdfunclclosed}, subject to the boundary condition \eqnref{currdef}.
To obtain Equation~\eqnref{airpdfunclclosed} from the fluid elements without droplets, one integrates their Fokker-Planck equation over initial positions. Equation~\eqnref{airpdfunclclosed} then follows, as a consequence of the relation between Lagrangian and Eulerian PDF:s mentioned above.

\subsection{Computation of observables and statistical one-dimensionality}
Since the PDF dynamics of our statistical model is implied by the dynamics of Lagrangian fluid elements, we solve Equations~\eqnref{pdfmodelclosed} for the PDF:s by evolving fluid elements according to Equations~\eqnref{velmcsuppl}, \eqnref{statmodsuppl} and \eqnref{radmcsuppl}.
The averages $\la s(\ve x,t)\ra$ and $\la r(t)s(\ve x,t)\ra$ are known in terms of the fluid elements, since the fluid elements sample $\mathscr{F}$ and $f$.
In practice, these averages are computed using kernel estimates, as described by \cite{pope2000turbulent}.
Since $\la s(\ve x,t)\ra$ and $\la r(t)s(\ve x,t)\ra$ are known, Equations~\eqnref{velmcsuppl}, \eqnref{statmodsuppl} and \eqnref{radmcsuppl} form a closed set of equations that govern the simultaneous evolution of an ensemble of fluid elements.
Since fluid elements with droplets sample $\mathscr{F}$, we can compute the droplet-size distribution and the fraction of completely evaporated droplets from them according to Equations~\eqnref{dsdstatmod} and \eqnref{pestatmod}.

To compute our results, we use the one-dimensional form of the statistical model shown in the main text.
This is possible, since the dynamics that we model is statistically one-dimensional.
Statistical one-dimensionality follows from that the turbulence is statistically homogeneous and isotropic, and from that our initial conditions are one-dimensional. The direction of statistical inhomogeneity is $x$. The PDF:s $\mathscr{F}(\ve U,S,R^2,\ve{x};t)$ and $f(\ve U,S;\ve x,t)$, as well as the averages  $\la r(t)s(\ve x,t) \ra$ and $\la s(\ve x,t) \ra$ that they form, depend on the three-dimensional position $\ve x$ only through $x$. As explained by \cite{pope1985pdf}, it is therefore sufficient to evolve only the $x$-components of fluid element positions and velocities. In the main text, we have taken the statistical one-dimensionality into account and replaced $\la r(t)s(\ve x,t) \ra$ and $\la s(\ve x,t) \ra$ by $\la r(t)s( x,t) \ra$ and $\la s( x,t) \ra$.

\subsection{Averaging}
The mean cubed radius of droplets and the volume average of supersaturation are required to compute the liquid-water potential temperature analogue $\theta$.
In the statistical model the mean cubed radius of droplets is given by:
\begin{align}\label{eq:r3statmod}\
\big{\langle} r(t)^3 \big{\rangle} = \frac{1}{ 1-P_\esubscript(t) } \int R^3 \mathscr{F}(\ve U,S,R^2,\ve{x};t) \,\mathrm d \ve U \mathrm d S \mathrm d R^2 \mathrm d \ve x\,.
\end{align}
The volume average of supersaturation is given by:
\begin{align}
\la s(t) \ra = \frac{1}{V} \int S f(\ve U,S;\ve{x},t)\, \mathrm d \ve U \mathrm d S \mathrm d \ve x\,.
\end{align}

\section{Simulation parameters}\label{sec:params}
The parameters of our simulations are listed in Table~\ref{tab:param}.
The parameters for Figures~2~to~4 are taken from the DNS of \cite{kumar2012extreme,kumar2014lagrangian,kumar2018scale}.
The results of \citep{kumar2014lagrangian,kumar2018scale} and \cite{Andrejczuk:2006} shown in these figures are not obtained using Equations~\maintextmicromodeqnnum, the microscopic equations that we model.
The interpretation of the simulation setups of \cite{kumar2014lagrangian,kumar2018scale} and \cite{Andrejczuk:2006} in terms of our parameters is discussed in Sections \ref{otherfundamental}~and~\ref{sec:DNS}.
\cite{kumar2018scale} did not give their values of $\profilefactor$, $\profileexponent$, and $s_\cloudairsubscript$, but we obtained them through private communication \citep{privcommkumar1}. For Figure~4, we use the values of $C_0$ and $L$ of Run 5 of \cite{kumar2018scale}, since we wanted to compare with DNS and are interested in large-Reynolds number regimes.
All simulations use $C_\phi = 2$, a common estimate for this empirical parameter \citep{pope2000turbulent}.

In Figure~5 we \eng{analyze}{analyse}  measurements from the top panel of Figure~3 of \cite{beals2015holographic}. In the supplemental material of that paper, the authors show droplet-size distributions and droplet-number densities of the undiluted cloud air.
The distribution of droplet radii that we use to analyse the top panel in Figure~3 of \cite{beals2015holographic} is very close to Gaussian, as explained in Section~\ref{sec:beals}. For our Figure~5, we fit a Gaussian to this distribution, and the fit is parameterised by the \dimensionless~standard deviation $\sigma_ 0 = 0.1386$.
We also use $C_0 = 6.5$ and $L = 2.28$ for Figure~5.
This value of $L$ is an estimate of the constant of proportionality between the \dimensionfull~domain size and $U\tau_L$ at large Reynolds numbers.
We compute this estimate by in two steps.
First, we estimate the turbulent kinetic energy associated to a spatial scale $\ell$ by integrating the energy-spectrum function \citep{pope2000turbulent} from the wave number $k = 2 \pi/\ell$:
\begin{align}\label{eq:kestimate}
\k = \int_{2 \pi/\ell}^\infty C \varepsilon^{\frac{2}{3}} k^{-\frac{5}{3}} \mathrm d k = C  \frac{3}{2}\lp \frac{\varepsilon \ell}{2 \pi} \rp^{\frac{2}{3}}.
\end{align}
Here, $C$ is a Kolmogorov constant that has been measured to $C = 1.5$ \citep{pope2000turbulent}.
Second, we use this value of the Kolmogorov constant and find
\begin{align}\label{eq:lestimate}
\frac{\ell}{U \tau_L} = \frac{4\pi}{3} C^{-\frac{3}{2}} = 2.28.
\end{align}
Since 2.28 is the constant of proportionality between an arbitrary spatial scale and $U\tau_L$, it is the constant of proportionality between the \dimensionfull~domain size and $U\tau_L$.
We therefore conclude that our estimate in Equation~\eqnref{kestimate} gives $L = 2.28$ and $V = L^3 = 11.85$.

Table~\ref{tab:param} provides a set of independent parameters that, together with $C_\phi = 2$, specify our simulations.
Also shown are a few dependent parameters. These include the Damk\"{o}hler number ratio $\ratio = \dad/\das$ and the critical Damk\"{o}hler number ratio
\begin{align}
\ratio_{\rm c} = -\frac{2}{3} \frac{\chi}{\left \langle s (0) \right \rangle},
\end{align}
where $\left \langle s (0) \right \rangle$ is the initial volume average of the supersaturation (defined below).
Also included is the coefficient  $\chi_0$ that determine $\left \langle s (0) \right \rangle$ together with $\chi$ and $s_\cloudairsubscript$.
The initial volume average of the supersaturation can be written
\begin{align}\label{eq:chizero}
\left \langle s (0) \right \rangle = (1+s_\cloudairsubscript)(\chi + \chi_0)-1,
\end{align}
a replacement that is done in the main text.
For Figure~5, we have $\chi_0 = s_\cloudairsubscript = 0$, so $\left \langle s (0) \right \rangle = \chi - 1$.
For Figures~2~to~4, we have a non-zero $\chi_0$, since the initial supersaturation profile is smooth.
Here, the parameters $\profilefactor$ and $\profileexponent$ that determine the initial supersaturation profile are tied to specific values of $\chi$.
It is only for these values of $\chi$ that the initial supersaturation is approximately zero at $x = \pm \chi L/2$, the edges of the cloud slab [Figure~1({\bf b})].
For a general $\chi$, the volume average of the initial supersaturation is given by Equation~\eqnref{chizero}, with a value of $\chi_0$ that can be computed from $\profilefactor$, $\profileexponent$, and $s_\cloudairsubscript$.

\section{Numerics}\label{sec:numparam}
We solve the statistical model using computer simulations that evolve an ensemble of Lagrangian fluid elements according to Equations~\maintextstatmodeqnnum.
Fluid elements with droplets are \eng{initialized}{initialised} uniformly over the interval $-\chi L/2 < x < \chi L/2$. To ensure that $\la s (x,t)\ra$ can be computed in the whole simulation domain, we \eng{initialize}{initialise} fluid elements without droplets uniformly over $-L/2 < x < L/2$ \citep{pope2000turbulent}.
We solve for the supersaturation $s(t)$ and squared radius $r^2(t)$ of fluid elements using Euler's method \citep{rade2013mathematics}.
We use a dynamical time step that ensures that the absolute errors in $s(t)$ and $r^2(t)$ at each time step are smaller than a given threshold.
At each time step, the position $x(t)$ and velocity $u(t)$ of a fluid element are drawn from their exact joint distribution, which is conditioned on the position and velocity before the time step \citep{gillespie1996exact}.
The averages $\la s(x,t) \ra$ and $\la r(t) s(x,t) \ra$ are computed from the fluid elements on a regular mesh.
We obtain the mean-field values at the positions of fluid elements by linear interpolation.
The spatial variations of these averages reduce with time due to mixing, and we increase the computational efficiency of our simulations by neglecting these spatial variations when they become very small.
That $\la s(x,t) \ra$ and $\la r(t) s(x,t) \ra$ become essentially independent of $x$ does not mean that the simulated system becomes spatially uniform.
In particular, individual \eng{realizations}{realisations} of the simulated system may still exhibit supersaturation fluctuations that affect the evaporation of droplets for some time.

Figures~2~and~3 show droplet-size distributions and time series of $P_e(t)$ for three simulations.
To conclude convergence for these simulations, we simply check that the results do not change as we vary the parameters and thresholds that control the numerics.
Figures~4~and~5 show results from hundreds of simulations.
To address the convergence of these simulations, we test the accuracy of our simulations for representative initial conditions and representative combinations of $\dad$, $\ratio$ and $\chi$. In each test, we vary the numerical parameters separately to exclude systematical errors. To estimate the statistical errors, we compute several independent \eng{realizations}{realisations} for each combination of the numerical parameters.
These tests allow us to conclude that no systematical errors pertain to the results in Figures~4~and~5, and that the values of $P_\esubscript^*$ that we compute have relative errors that are less than 5 \%, and/or absolute errors that are smaller than $10^{-3}$. The simulation results in Figures~4~and~5  are either $P_\esubscript^*$ or functions of $P_\esubscript^*$, and we conclude that none of our conclusions in the main text are affected by numerical errors.

\section{Parameters of DNS in Figure 4}\label{sec:DNS}

To place DNS of \cite{Andrejczuk:2006,kumar2012extreme,Kumar:2013,kumar2014lagrangian,kumar2018scale} in our phase diagram (Figure~4), we extract their \dimensionfull~ parameters that determine $\dad$ and $\ratio/\ratio_{\rm c}$.
The computed values are listed in Table~\tabref{dnskumar} for \cite{kumar2012extreme,Kumar:2013,kumar2014lagrangian,kumar2018scale}, and in Table~\tabref{dnsandr} for \cite{Andrejczuk:2006}.

\subsection{DNS of \cite{kumar2012extreme,Kumar:2013,kumar2014lagrangian,kumar2018scale}}
It is straightforward to compute $\dad$, $\ratio$, and $\ratio_{\rm c}$ for the simulations in \cite{kumar2012extreme,Kumar:2013,kumar2018scale} and three simulations in \cite{kumar2014lagrangian}, since they are for stationary turbulence with Lagrangian droplets, just as Equations~1.
First, one casts the dynamics into the dynamics of Equations~\eqnref{modelsimplified}, using the simplifications  described in Section~\ref{otherfundamental}.
After that one \dedimensionalise~as described in Section~\ref{sec:dimensionless}.
The simulations of \cite{kumar2012extreme,Kumar:2013,kumar2014lagrangian,kumar2018scale} shown in Figure~4 are the simulations for which one can conclude if $\pe^* > 10 \%$ or if $\pe^* < 10 \%$.

\subsection{DNS of \cite{Andrejczuk:2006}}
Figure~4 shows 42 simulations of \cite{Andrejczuk:2006}. \cite{Andrejczuk:2006} reports of 58 simulations, but we can only read off whether $\pe^* > 10 \%$ or not for 50 of them.
Out of these 50 simulations, eight simulations fall outside the plot range of Figure~4. We therefore show 42 simulations of \cite{Andrejczuk:2006} in Figure~4.

The simulations of \cite{Andrejczuk:2006} are for decaying turbulence and initial conditions in the form of randomly distributed cloudy filaments in dry air.
The parameter $\dad$ in Equation~\eqnref{das} is defined for the stationary turbulence and the initial conditions of our simulations, so it has no direct counterpart in \cite{Andrejczuk:2006}.
To nevertheless discuss the simulations of \cite{Andrejczuk:2006} we compute a time-scale ratio that incorporates the same physics as $\dad$.
We then place a simulation in Figure~4 by assuming that this time-scale ratio is equal to $\dad$.
The time-scale ratio is computed as $\dad = \tau_L/\tau_{\rm d}$, but with $\tau_L$ taken as $\k/\varepsilon$ at the time when $\k$ decayed fastest in \cite{Andrejczuk:2006}.
We estimate these values of $\k$ and $\varepsilon$ from plots in \cite{Andrejczuk:2006}.

The coefficient $A_3$ enters the dynamics multiplied by an unnamed factor $\gamma_{A_3}$ in \cite{Andrejczuk:2006}, and we read off $A_3/\gamma_{A_3} = 10^{-10}$ m$^2$/s.
We extract the the density $\varrho_a$ of air and the density $\varrho_{\rm w}$ of pure liquid water \cite{Andrejczuk:2006,andrejczuk2004numerical}.
The expression for saturation pressure of water vapour is not given in \cite{Andrejczuk:2006}, so we can not compute their value of $A_2$ using Equation~\eqnref{a2other}.
We therefore use the value $A_2 = 413$~m$^3$/kg that we compute for \cite{kumar2014lagrangian}.
To nevertheless analyse DNS results of \cite{Andrejczuk:2006} qualitatively is justified since the temperatures and water-vapour densities in \cite{Andrejczuk:2006} and \cite{kumar2014lagrangian} imply that thair values of $A_2$ are similar.
We compute $s_e$ as one minus the relative humidity of the non-cloudy initial filaments, which we extract from tables in  \cite{Andrejczuk:2006}.
We also extract the volume fraction $\chi$ of cloudy air, and the liquid-water mixing ratio $q_\ell$ of the initially cloudy filaments from these tables.
We interpret the initially cloudy filaments as saturated and possesing sharp edges, even though this is not stated explicitly in \cite{Andrejczuk:2006}, so that $\ratio_c = -2 \chi/[3(\chi -1)]$.

\cite{Andrejczuk:2006} do not use Lagrangian droplets. Instead, they employ a field description in which droplets are binned into 16 size categories.
The three initially populated bins are  centered at droplet radii  $8$, $8.75$, and $9.5$ $\mu$m, and they contained 25\%, 50\%, and 25\% of the liquid-water mixing ratio $q_\ell$.
We compute the number density of droplets within a bin centered at the droplet radius $r$ that contains a fraction $\xi_r$ of liquid-water mixing ratio as $n_r = 3 \xi_r q_\ell \varrho_a/(4 \pi r^3 \varrho_{\rm w})$. By summing up the number densities for $r=8$, $8.75$, and $9.5$ $\mu$m, we find the droplet-number density $n_0$ of the initially cloudy filaments.
We compute the initial volume radii of droplets as $r_0 = [ 3 q_\ell \varrho_a/(4 \pi n_0 \varrho_{\rm w})]^{1/3}$

\section{Estimates used to Analyse data from \cite{beals2015holographic}}\label{sec:beals}
The black crosses in Figure~5{\bf(a)} are from the top panel of Figure~3 of \cite{beals2015holographic}, and represent droplets measured during a research flight through a convective cloud. Droplet-size distributions and thermodynamic conditions are not given for this flight, so we estimate them using data from a typical flight detailed in the supplementary materials of \cite{beals2015holographic}, namely pass 2 of the research flight RF05.

To estimate the parameters of our model, we need actual values for physical coefficients.
We use $\Lambda = 2.5\cdot 10^6$ J/kg, $c_p = 1005$ J/(kg$\cdot$K), and $R_{\rm v} = 461.5$ J/(kg$\cdot$K) from \cite{rogers1989short}. We also set the density of pure liquid water to $\varrho_{\rm w} = 1000$~kg/m$^3$.
Figure~S3 of \cite{beals2015holographic} shows that pass 2 of RF05 traverses a cloud at a constant altitude of 4000 m.
We therefore assume the pressure of the International Standard Atmosphere \citep{cavcar2000international} at this altitude, $p \sim 62000$~Pa.
The precise value of the pressure is not important, since conclusions that rely on the estimates in this Section are unchanged if we assume a pressure that is 10 \% lower or higher.
We extract $K_{\rm a}$ and $D_{\rm v}$ at our assumed pressure from Table 7.1 of  \cite{rogers1989short}.
Furthermore, we assume that the saturation pressure of water vapour is given by Equation~(2.12) in \cite{rogers1989short}.
We also use the relation between potential temperature $\Theta$, temperature $T$, and pressure $p$ of \cite{rogers1989short}, $\Theta = T(10^5\,{\rm Pa}/p)^{0.286}$.

The droplet-size distribution of undiluted cloud air traversed during pass 2 of RF05 is shown in Figure~S7 of \cite{beals2015holographic}. The distribution of droplet radii is very close Gaussian, normalised to the number-density of droplets.
We make a Gaussian fit with mean $4.42$ $\mu$m, and standard deviation $0.626$ $\mu$m. The corresponding volume radius is $r_0 = 4.51$ $\mu$m, which gives the \dimensionless~ standard deviation $\sigma_ 0 = 0.1386$.
We use this value of $\sigma_ 0$ to parameterise the initial droplet-size distribution in the statistial-model simulations shown in Figure~5.
The normalisation of the Gaussian fit gives the droplet-number density $n_0 = 764$ cm$^{-3}$.
We compute the liquid-water density of the undiluted cloud as $\varrho_{\ell \rm c}= (4 \pi r_0^3/3)n_0 \varrho_{\rm w} = 2.94\cdot10^{-4}$~kg/m$^3$. This value of $\varrho_{\ell \rm c}$ is consistent with the liquid-water density plot in Figure~S4 of \cite{beals2015holographic}.

We estimate temperatures from the liquid-water potential temperature shown in Figure~S4 of \cite{beals2015holographic}.
Here, the liquid-water potential temperature varies between $\Theta_{\ell \rm c} = 294.8$~K within the cloud and $\Theta_{\ell \rm e} = 295.8$ K outside the cloud.
The liquid-water potential temperature is the same as the potential temperature $\Theta_{\rm e}$ outside the cloud, $\Theta_{\rm e} = 295.8$~K.
We compute the temperature $T_{\rm e} \sim 257.6$ outside the cloud, using $\Theta_{\rm e}$ and the pressure of the International Standard Atmosphere \citep{cavcar2000international}.
We estimate the density of dry air as $\airdensity  = p/(R_{\rm v} T_{\rm e}) \sim 0.52$ kg/m$^3$. This density gives the liquid-water mixing ratio $q_{\ell \rm c} = \varrho_{\ell \rm c}/\airdensity \sim 5.6\cdot10^{-4}$ within the cloud.
The potential temperature within the cloud can now be estimated as $\Theta_{\rm c} = \Theta_{\ell \rm c} + (\Lambda/c_p)q_{\ell \rm c} \sim 296.2$~K \citep{lamb2011physics}, and the corresponding temperature estimate is $T_{\rm c} \sim 257.9$~K.
It is seen in Figure~S4 of \cite{beals2015holographic} that the supersaturation within the cloud is smaller than $\sim$2\% in magnitude. Assuming saturation within the cloud, the estimate $T_{\rm c} \sim 257.9$~K gives the water-vapour mixing ratio $q_{\rm vc} = e_{\rm vs}(T_{\rm c})/(\airdensity R_{\rm v}T_{\rm c})\sim 3.1\cdot10^{-3}$. We extract the (negative) supersaturaion $s_{\rm e} \sim -0.08$ from Figure~S4 of \cite{beals2015holographic}, and estimate the water-vapour mixing ratio outside the cloud, $q_{\rm ve} = e_{\rm vs}(T_{\rm e})(s_{\rm e}+1)/(\airdensity R_{\rm v}T_{\rm e})\sim 2.7\cdot10^{-3}$.

In our analysis of empirical data from \cite{beals2015holographic} we estimate $\tau_{\rm s} \sim 1$~s based on our estimates of $T_{\rm c}$, $q_{\rm vc}$, and $p$.
Equation~\eqnref{a2other} gives $A_2 \sim 1000$~$m^3$/kg when linearising the supersaturation around the temperature $T_{\rm c}\sim 257.9$~K and the water-vapour mixing ratio $q_{\rm vc} \sim 3.1\cdot10^{-3}$. We compute $A_3\sim 2\cdot10^{-11}$~m$^2$/s at the pressure $p = 62000$~Pa and temperature $T_{\rm c}\sim 257.9$~K using Equation~\eqnref{athreedef}. Using the volume radius $r_0$ and droplet-number density $n_0$ estimated above, we find $\tau_{\rm s} \sim 1$~s using Equation~\eqnref{taus}.

\newpage

\begin{landscape}
\part*{Table S1}

\begin{table}[h]
\centering
\captionsetup{width=160mm}
\footnotesize
  \caption{\label{tab:param} Simulation parameters.
Our statistical-model simulations are completely specified by the two Damk\"{o}hler numbers $ \normalfont \dad$ and $\normalfont \das$, the constant $C_0$ that regulates the auto-correlation time of fluid elements, the domain size $L$, the volume fraction $\chi$ of cloudy air, the constants $\profilefactor$ and $\profileexponent$ that determine the shape of the initial supersaturation profile in Figures~2~to~4, and the supersaturation $s_\cloudairsubscript$ at the center of the initial cloud slab, together with the empirical constant $C_\phi = 2$.
From these parameters, we compute the Damk\"{o}hler number ratio $\ratio$, the critical Damk\"{o}hler number ratio $\ratio_{\rm c}$, and the contribution $\chi_0$ to the initial volume average of the supersaturation from the shape of the initial supersaturation profile.
}
\begin{tabular}{l|lllllllll|lll}
\hline \hline
\multirow{2}{*}{Simulation} & \multicolumn{9}{c|}{  Independent parameters}                             & \multicolumn{3}{l}{ Dependent parameters}  \\
                            & $\dad$    & $\das$    & $C_0$   & $L$    & $\chi$    & $\sigma_0$ & $\profilefactor$ & $\profileexponent$ & $s_\cloudairsubscript$    & $\ratio$            & $\ratio_{\rm c}$          & $\chi_0$                    \\ \hline
Figures~2 and 3, dry             & 2.44   & 0.968 & 5.22 & 2.96 & 0.428  & 0      & 4722  & 8     & 0.021 & 2.52         & 0.859      & 0.226          \\
Figure~2, moist             & 1.09   & 1.433 & 5.22 & 2.96 & 0.428  & 0      & 4722  & 8     & 0.021 & 0.760         & 0.859      & 0.226           \\
Figure~3, very moist                   & 0.754 & 8.2    & 4.50 & 2.99 & 0.4    & 0      & 1410  & 6     & 0.1   & 0.092        & 0.683      & 0.154      \\
Figure~4                & 5E$^{\text{-3}}$-4E$^{\text{2}}$ & 1E$^{\text{-3}}$-9E$^{\text{3}}$ & 6.09  & 2.66 & 0.429  & 0      & 690   & 6     & 0.1   & 5E$^{\text{-2}}$-4E$^{\text{0}}$      & 0.913      & 0.195            \\
Figure~5({\bf a})              & 1E$^{\text{-2}}$-1E$^{\text{3}}$ & 6E$^{\text{-2}}$-6E$^{\text{3}}$  & 6.5  & 2.28 &0.2-0.8 & 0.1386 &      &      & 0     & 0.17         & 0.17-2.7  & 0             \\
Figure~5({\bf b})           & 1E$^{\text{-2}}$-1E$^{\text{3}}$  & 3E$^{\text{-1}}$-4E$^{\text{4}}$ & 6.5  & 2.28 & 0.369-0.374 & 0.1386 &      &      & 0     & 2E$^{\text{-2}}$-3E$^{\text{-2}}$     & 0.38-0.41      & 0                  \\ \hline      
\end{tabular}
\end{table}
\end{landscape}

\newpage
{\huge \bf Table S2}
\begin{table}[h!]
\small
\centering
\setlength{\tabcolsep}{6pt}
\caption{\label{tab:dnskumar} Same as Table~2, but separately for each DNS from \cite{kumar2012extreme,Kumar:2013,kumar2014lagrangian,kumar2018scale} in Figure~4. \Dimensionless~parameters: Damk\"{o}hler number $\dad$, Damk\"{o}hler-number ratio $\ratio$, critical ratio $\ratio_{\rm c}$, and volume fraction $\chi$ of cloudy air. \Dimensionfull~ parameters: domain size $L$, mean dissipation rate $\varepsilon$, and droplet-number density $n_0$ of the initially cloudy air.
}
\begin{threeparttable}
\begin{tabular}{lllllllll}
\headrow
  &  \multicolumn{5}{c}{\bf \Dimensionless~parameters} & \multicolumn{3}{c}{\bf \Dimensionfull~parameters} \\[-1.0ex]
\headrow & & & & & & $\,\,\,L$&   $\,\,\,\,\,\,\,\varepsilon$ & $\,\,\,\,\,n_0$ \\[-0.9ex]
\headrow
\multirow{-1.5}{*}{\bf Reference}& \multirow{-1.5}{*}{$\dad$} & \multirow{-1.5}{*}{$\das$} &  \multirow{-1.5}{*}{$\ratio$}& \multirow{-1.5}{*}{$\ratio_{\rm c}$}  & \multirow{-1.5}{*}{$\chi$}  & [cm] &  [cm$^2$/s$^3$] &   [cm$^{-3}$]  \\
\hiderowcolors
\multicolumn{1}{l|}{Table 2 in \cite{kumar2012extreme}, Row 1}&\multicolumn{1}{|l}{0.0075}&0.0820&0.0916&0.6829&\multicolumn{1}{l|}{ }&\multicolumn{1}{|l}{25.6}&33.8&164\\  
\multicolumn{1}{l|}{Table 2 in \cite{kumar2012extreme}, Row 2}&\multicolumn{1}{|l}{0.0751}&0.8204&0.0916&0.6829&\multicolumn{1}{l|}{ }&\multicolumn{1}{|l}{25.6}&33.8&164\\  
\multicolumn{1}{l|}{Table 2 in \cite{kumar2012extreme}, Row 3}&\multicolumn{1}{|l}{0.7511}&8.2041&0.0916&0.6829&\multicolumn{1}{l|}{ }&\multicolumn{1}{|l}{25.6}&33.8&164\\  
\multicolumn{1}{l|}{Table 2 in \cite{Kumar:2013}, Row 3}&\multicolumn{1}{|l}{0.3065}&0.4185&0.7324&0.6829&\multicolumn{1}{l|}{ }&\multicolumn{1}{|l}{25.6}&33.8&164\\        
\multicolumn{1}{l|}{Table 2 in \cite{Kumar:2013}, Row 7}&\multicolumn{1}{|l}{0.1362}&0.6277&0.2170&0.6829&\multicolumn{1}{l|}{ }&\multicolumn{1}{|l}{25.6}&33.8&164\\        
\multicolumn{1}{l|}{Simulation S1 in \cite{kumar2014lagrangian}}&\multicolumn{1}{|l}{2.4320}&0.9660&2.5177&0.8377&\multicolumn{1}{l|}{ }&\multicolumn{1}{|l}{51.2}&33.8&153\\
\multicolumn{1}{l|}{Simulation S2 in \cite{kumar2014lagrangian}}&\multicolumn{1}{|l}{1.0809}&1.4490&0.7460&0.8377&\multicolumn{1}{l|}{ }&\multicolumn{1}{|l}{51.2}&33.8&153\\
\multicolumn{1}{l|}{Simulation S3 in \cite{kumar2014lagrangian}}&\multicolumn{1}{|l}{0.6080}&1.9319&0.3147&0.8377&\multicolumn{1}{l|}{ }&\multicolumn{1}{|l}{51.2}&33.8&153\\
\multicolumn{1}{l|}{Run 1 in \cite{kumar2018scale}}&\multicolumn{1}{|l}{0.1191}&0.5185&0.2296&0.9150&\multicolumn{1}{l|}{ }&\multicolumn{1}{|l}{12.8}&31.9&118\\             
\multicolumn{1}{l|}{Run 2 in \cite{kumar2018scale}}&\multicolumn{1}{|l}{0.1914}&0.8333&0.2297&0.9155&\multicolumn{1}{l|}{ }&\multicolumn{1}{|l}{25.6}&34.6&118\\             
\multicolumn{1}{l|}{Run 3 in \cite{kumar2018scale}}&\multicolumn{1}{|l}{0.3227}&1.4042&0.2298&0.9163&\multicolumn{1}{l|}{ }&\multicolumn{1}{|l}{51.2}&34.7&118\\             
\multicolumn{1}{l|}{Run 4 in \cite{kumar2018scale}}&\multicolumn{1}{|l}{0.5842}&2.4509&0.2384&0.9503&\multicolumn{1}{l|}{ }&\multicolumn{1}{|l}{102.4}&32.1&113\\            
\multicolumn{1}{l|}{Run 5 in \cite{kumar2018scale}}&\multicolumn{1}{|l}{0.9122}&4.0429&0.2256&0.9000&\multicolumn{1}{l|}{ }&\multicolumn{1}{|l}{204.8}&33.6&120\\      
\hline
\end{tabular}
\end{threeparttable}
\end{table}

\clearpage
{\huge \bf Table S3}
\begin{table}[h!]
\small
\centering
\setlength{\tabcolsep}{10pt}
\caption{\label{tab:dnsandr} Same as Table~2, but separately for each DNS from \cite{Andrejczuk:2006} in Figure~4. \Dimensionless~parameters: Damk\"{o}hler number $\dad$, Damk\"{o}hler-number ratio $\ratio$, critical ratio $\ratio_{\rm c}$, and volume fraction $\chi$ of cloudy air. \Dimensionfull~parameters: domain size $L$, mean dissipation rate $\varepsilon$, and droplet-number density $n_0$ of the initially cloudy air.
}
\begin{threeparttable}
\def\arraystretch{0.89}
\begin{tabular}{lllllllll}
\headrow
  &  \multicolumn{5}{c}{\bf \Dimensionless~parameters} & \multicolumn{3}{c}{\bf \Dimensionfull~parameters} \\[0ex]
\headrow & & & & & & $\,\,\,L$&   $\,\,\,\,\,\,\,\varepsilon$ & $\,\,\,\,\,n_0$ \\[0ex]
\headrow
\multirow{-1.5}{*}{\bf \#}& \multirow{-1.5}{*}{$\dad$} & \multirow{-1.5}{*}{$\das$} &  \multirow{-1.5}{*}{$\ratio$}& \multirow{-1.5}{*}{$\ratio_{\rm c}$}  & \multirow{-1.5}{*}{$\chi$}  & [cm] &  [cm$^2$/s$^3$] &   [cm$^{-3}$]  \\
\hiderowcolors
\multicolumn{1}{l|}{S2a$_1$}&\multicolumn{1}{|l}{13.7874}&78.1243&0.1765&0.0996&\multicolumn{1}{l|}{0.13}&\multicolumn{1}{|l}{64}&1.5&1166\\
\multicolumn{1}{l|}{S2a$_2$}&\multicolumn{1}{|l}{6.1722}&34.9739&0.1765&0.3284&\multicolumn{1}{l|}{0.33}&\multicolumn{1}{|l}{64}&3&1166\\   
\multicolumn{1}{l|}{S2a$_3$}&\multicolumn{1}{|l}{12.1003}&68.5642&0.1765&0.5029&\multicolumn{1}{l|}{0.43}&\multicolumn{1}{|l}{64}&1.7&1166\\
\multicolumn{1}{l|}{S2a$_4$}&\multicolumn{1}{|l}{9.6240}&54.5331&0.1765&0.6667&\multicolumn{1}{l|}{0.5}&\multicolumn{1}{|l}{64}&1.7&1166\\  
\multicolumn{1}{l|}{S2a$_5$}&\multicolumn{1}{|l}{10.1375}&57.4425&0.1765&1.3535&\multicolumn{1}{l|}{0.67}&\multicolumn{1}{|l}{64}&1.5&1166\\
\multicolumn{1}{l|}{S2b$_1$}&\multicolumn{1}{|l}{4.3101}&24.4223&0.1765&0.0996&\multicolumn{1}{l|}{0.13}&\multicolumn{1}{|l}{64}&8.1&1166\\ 
\multicolumn{1}{l|}{S2b$_2$}&\multicolumn{1}{|l}{4.3101}&24.4223&0.1765&0.3284&\multicolumn{1}{l|}{0.33}&\multicolumn{1}{|l}{64}&8.1&1166\\ 
\multicolumn{1}{l|}{S2b$_3$}&\multicolumn{1}{|l}{4.3101}&24.4223&0.1765&0.5029&\multicolumn{1}{l|}{0.43}&\multicolumn{1}{|l}{64}&8.1&1166\\ 
\multicolumn{1}{l|}{S2b$_4$}&\multicolumn{1}{|l}{4.3101}&24.4223&0.1765&0.6667&\multicolumn{1}{l|}{0.5}&\multicolumn{1}{|l}{64}&8.1&1166\\  
\multicolumn{1}{l|}{S2b$_5$}&\multicolumn{1}{|l}{4.3101}&24.4223&0.1765&1.3535&\multicolumn{1}{l|}{0.67}&\multicolumn{1}{|l}{64}&8.1&1166\\ 
\multicolumn{1}{l|}{S2c$_1$}&\multicolumn{1}{|l}{0.8232}&4.6646&0.1765&0.0996&\multicolumn{1}{l|}{0.13}&\multicolumn{1}{|l}{64}&935.3&1166\\
\multicolumn{1}{l|}{S2c$_2$}&\multicolumn{1}{|l}{0.8232}&4.6646&0.1765&0.3284&\multicolumn{1}{l|}{0.33}&\multicolumn{1}{|l}{64}&935.3&1166\\
\multicolumn{1}{l|}{S2c$_3$}&\multicolumn{1}{|l}{0.8232}&4.6646&0.1765&0.6667&\multicolumn{1}{l|}{0.5}&\multicolumn{1}{|l}{64}&935.3&1166\\ 
\multicolumn{1}{l|}{S3a$_2$}&\multicolumn{1}{|l}{8.2897}&7.3394&1.1295&0.6667&\multicolumn{1}{l|}{0.5}&\multicolumn{1}{|l}{64}&1.3&182\\    
\multicolumn{1}{l|}{S3a$_3$}&\multicolumn{1}{|l}{11.6711}&20.6664&0.5647&0.6667&\multicolumn{1}{l|}{0.5}&\multicolumn{1}{|l}{64}&0.9&364\\  
\multicolumn{1}{l|}{S3a$_4$}&\multicolumn{1}{|l}{9.7721}&25.9556&0.3765&0.6667&\multicolumn{1}{l|}{0.5}&\multicolumn{1}{|l}{64}&1.4&547\\   
\multicolumn{1}{l|}{S3a$_5$}&\multicolumn{1}{|l}{8.4295}&47.7646&0.1765&0.6667&\multicolumn{1}{l|}{0.5}&\multicolumn{1}{|l}{64}&2&1166\\    
\multicolumn{1}{l|}{S4a$_1$}&\multicolumn{1}{|l}{18.0709}&47.9980&0.3765&0.0996&\multicolumn{1}{l|}{0.13}&\multicolumn{1}{|l}{64}&0.5&547\\ 
\multicolumn{1}{l|}{S4a$_2$}&\multicolumn{1}{|l}{7.3957}&19.6436&0.3765&0.3284&\multicolumn{1}{l|}{0.33}&\multicolumn{1}{|l}{64}&1.6&547\\  
\multicolumn{1}{l|}{S4a$_3$}&\multicolumn{1}{|l}{8.4955}&22.5650&0.3765&0.5029&\multicolumn{1}{l|}{0.43}&\multicolumn{1}{|l}{64}&1.5&547\\  
\multicolumn{1}{l|}{S4a$_4$}&\multicolumn{1}{|l}{9.7754}&25.9643&0.3765&0.6667&\multicolumn{1}{l|}{0.5}&\multicolumn{1}{|l}{64}&1.3&547\\   
\multicolumn{1}{l|}{S4a$_5$}&\multicolumn{1}{|l}{12.6777}&33.6732&0.3765&1.3535&\multicolumn{1}{l|}{0.67}&\multicolumn{1}{|l}{64}&1.3&547\\ 
\multicolumn{1}{l|}{S4b$_2$}&\multicolumn{1}{|l}{14.1203}&12.5016&1.1295&0.3284&\multicolumn{1}{l|}{0.33}&\multicolumn{1}{|l}{64}&0.4&182\\ 
\multicolumn{1}{l|}{S4b$_3$}&\multicolumn{1}{|l}{13.1643}&11.6552&1.1295&0.5029&\multicolumn{1}{l|}{0.43}&\multicolumn{1}{|l}{64}&0.6&182\\ 
\multicolumn{1}{l|}{S4b$_4$}&\multicolumn{1}{|l}{8.1755}&7.2383&1.1295&0.6667&\multicolumn{1}{l|}{0.5}&\multicolumn{1}{|l}{64}&1.1&182\\    
\multicolumn{1}{l|}{S4b$_5$}&\multicolumn{1}{|l}{9.8085}&8.6841&1.1295&1.3535&\multicolumn{1}{l|}{0.67}&\multicolumn{1}{|l}{64}&1.2&182\\   
\multicolumn{1}{l|}{S4b$_6$}&\multicolumn{1}{|l}{13.9764}&12.3742&1.1295&4.4615&\multicolumn{1}{l|}{0.87}&\multicolumn{1}{|l}{64}&0.9&182\\ 
\multicolumn{1}{l|}{S4c$_5$}&\multicolumn{1}{|l}{9.6947}&3.4333&2.8237&1.3535&\multicolumn{1}{l|}{0.67}&\multicolumn{1}{|l}{64}&1&73\\      
\multicolumn{1}{l|}{S4c$_6$}&\multicolumn{1}{|l}{16.1282}&5.7117&2.8237&4.4615&\multicolumn{1}{l|}{0.87}&\multicolumn{1}{|l}{64}&0.6&73\\   
\multicolumn{1}{l|}{S5a$_1$}&\multicolumn{1}{|l}{12.0940}&34.2643&0.3530&0.6667&\multicolumn{1}{l|}{0.5}&\multicolumn{1}{|l}{64}&3.7&1166\\ 
\multicolumn{1}{l|}{S5a$_2$}&\multicolumn{1}{|l}{11.6186}&41.8950&0.2773&0.6667&\multicolumn{1}{l|}{0.5}&\multicolumn{1}{|l}{64}&3.1&1166\\ 
\multicolumn{1}{l|}{S5a$_3$}&\multicolumn{1}{|l}{9.3465}&52.9606&0.1765&0.6667&\multicolumn{1}{l|}{0.5}&\multicolumn{1}{|l}{64}&1.7&1166\\  
\multicolumn{1}{l|}{S5a$_4$}&\multicolumn{1}{|l}{7.7943}&61.8316&0.1261&0.6667&\multicolumn{1}{l|}{0.5}&\multicolumn{1}{|l}{64}&1.1&1166\\  
\multicolumn{1}{l|}{S6a$_1$}&\multicolumn{1}{|l}{13.3405}&37.7959&0.3530&0.6667&\multicolumn{1}{l|}{0.5}&\multicolumn{1}{|l}{64}&2.7&1166\\ 
\multicolumn{1}{l|}{S6a$_2$}&\multicolumn{1}{|l}{14.7497}&41.7886&0.3530&0.6667&\multicolumn{1}{l|}{0.5}&\multicolumn{1}{|l}{64}&3.4&1166\\ 
\multicolumn{1}{l|}{S6a$_3$}&\multicolumn{1}{|l}{12.8279}&36.3436&0.3530&0.6667&\multicolumn{1}{l|}{0.5}&\multicolumn{1}{|l}{64}&5.9&1166\\ 
\multicolumn{1}{l|}{S6a$_4$}&\multicolumn{1}{|l}{10.1435}&28.7383&0.3530&0.6667&\multicolumn{1}{l|}{0.5}&\multicolumn{1}{|l}{64}&10.2&1166\\
\multicolumn{1}{l|}{S6b$_1$}&\multicolumn{1}{|l}{1.6879}&4.7821&0.3530&0.6667&\multicolumn{1}{l|}{0.5}&\multicolumn{1}{|l}{64}&3&1166\\     
\multicolumn{1}{l|}{S6b$_2$}&\multicolumn{1}{|l}{3.5983}&10.1947&0.3530&0.6667&\multicolumn{1}{l|}{0.5}&\multicolumn{1}{|l}{64}&3.1&1166\\  
\multicolumn{1}{l|}{S6b$_3$}&\multicolumn{1}{|l}{13.1067}&37.1334&0.3530&0.6667&\multicolumn{1}{l|}{0.5}&\multicolumn{1}{|l}{64}&3.8&1166\\ 
\multicolumn{1}{l|}{S6b$_4$}&\multicolumn{1}{|l}{56.6502}&160.4998&0.3530&0.6667&\multicolumn{1}{l|}{0.5}&\multicolumn{1}{|l}{64}&3.4&1166\\
\multicolumn{1}{l|}{S6b$_5$}&\multicolumn{1}{|l}{97.6054}&276.5329&0.3530&0.6667&\multicolumn{1}{l|}{0.5}&\multicolumn{1}{|l}{64}&4.7&1166\\
\hline
\end{tabular}
\end{threeparttable}
\end{table}

\clearpage
\part*{SI References}
\begingroup
\renewcommand{\section}[2]{}

\endgroup